\documentclass[runningheads]{llncs}
\usepackage{graphicx}
\usepackage{amsmath,amssymb} 
\DeclareMathOperator*{\argmax}{arg\,max}
\DeclareMathOperator*{\argmin}{arg\,min}
\usepackage{color}

\usepackage{tabularx}
\usepackage{hyperref}
\usepackage{bbm}
\graphicspath{{images/}, {images/teaser/}, {images/pipeline/}, {images/GAN_results_2D_textures/}, {images/3DMM_2d_renders/}, {images/3DMM_distribution/},
{images/Geom_syn/},
{images/tsne/},
{images/NN/},
{images/NN-new/},
{images/NN-experiment/},
{images/3DMM_low_high_freqs/}}
\usepackage{caption}
\usepackage[position=b]{subcaption}
\begin{document}

\title{High Quality Facial Surface and Texture Synthesis
   via Generative Adversarial Networks} 
   \titlerunning{Facial geometry synthesis by GAN}
   \author{Ron Slossberg\inst{1}* \and
   Gil Shamai\inst{2}* \and
   Ron Kimmel\inst{1} }

   %

   \institute{Technion CS department, Haifa,Israel \and
   Technion EE departemnt, Haifa, Israel
   \\
   \email{\{ronslos,sgils\}@campus.technion.ac.il}\\
   \email{ron@cs.technion.ac.il}\\
   * Indicates equal contribution by authors}

\maketitle

\begin{abstract}
In the past several decades, many attempts have been made to model synthetic realistic geometric data.
The goal of such models is to generate plausible 3D geometries and textures. Perhaps the best known of its kind is the linear 3D morphable model (3DMM) for faces.
Such models can be found at the core of many computer vision applications such as face reconstruction, recognition and authentication to name just a few.

Generative adversarial networks (GANs) have shown great promise in imitating high dimensional data distributions.
State of the art GANs are capable of performing tasks such as image to image translation as well as auditory and image signal synthesis, producing novel plausible samples from the data distribution at hand.

Geometric data is generally more difficult to process due to the inherent lack of an intrinsic parametrization.
By bringing geometric data into an aligned space, we are able to map the data onto a 2D plane using a universal parametrization. This alignment process allows for efficient processing of digitally scanned geometric data via image processing tools.

Using this methodology, we propose a novel face synthesis model for generation of realistic facial textures together with their corresponding geometry. A GAN is employed in order to imitate the space of parametrized human textures, while corresponding facial geometries are generated by learning the best 3DMM coefficients for each texture.
The generated textures are mapped back onto the corresponding geometries to obtain new generated high resolution 3D faces.

\end{abstract}

\section{Introduction}
In recent years, deep learning has gained popularity in many research fields as well as practical applications.
Deep networks are powerful generalization tools which are able to answer complex questions about data samples in a surprisingly effective manner.
It has been well established that in order to train highly complex models, it is necessary to obtain extensive amounts of training data which closely approximates the complete data distribution.

Data augmentation is a popular method for extending the size of a given dataset. The idea is to modify the training examples in such a way that keeps their semantic properties intact. For instance, one can apply basic geometric distortions or add noise to a photo of an object in a way that leaves the object recognizable. Though helpful in many cases, these simple data augmentation methodologies often fail to address more complex transformations of the data such as pose, lighting and non-rigid deformations. An example of a more advanced type of data augmentation is demonstrated by \cite{masi2016we}, who observed that augmenting facial data by applying geometric and photometric transformations increases the performance of facial recognition models.

A different trend in data acquisition and augmentation for training deep networks is to synthesize training examples using a simulator such as \cite{shrivastava2017learning}.
The simulator should be able to model and generate a rich variety of samples which can be constructed under controlled conditions such as pose and lighting. However, synthetically generated examples often look unrealistic and diverge from the distribution of natural data.
Methods such as \cite{richardson2017learning} that used unrealistic synthetic data for training their models had to contend with difficulties when applying their models onto real data. A more realistic simulator that captures the real world data statistics more accurately would be expected to allow for easier generalization to real data.

In this line of works, recent papers have focused on making synthetic data more realistic by using {\em generative adversarial networks} (GANs).
Commonly, the simulated data is used as an input to the GAN which can produce a more realistic sample from the synthetic one \cite{gecer2018semi,shrivastava2017learning}.
Taking this approach, the generated samples may appear realistic, however their semantic properties might be altered during the process, even when imposing a loss which penalizes the change in the parameters of the output.

Reducing the scope to modeling photo-geometric data, one of the most commonly used models for representation and synthesis of geometries and textures is the 3DMM \cite{blanz1999morphable} (see \autoref{sec:3dmm}), originally proposed in the context of 3D human faces. Using a simple linear representation, 3DMM is capable of providing various new samples of the data it is constructed from. However, the generated samples are unrealistic in appearance and since the generation model follows a Gaussian distribution, non-plausible samples may easily result from the generation process.

Here, we propose a new realistic data synthesis approach for human faces.
The suggested approach does not suffer from indirect control over various desired attributes such as pose and lighting, yet still produces realistic looking plausible models, in contrast to \cite{blanz1999morphable}.
Moreover, in contrast to \cite{masi2016we,saito2017photorealistic} the proposed model is not limited to producing new instances of existing individuals, but instead is capable of generating new plausible identities. This synthesis would be beneficial for various applications such as face recognition and biometric verification, as well as face reconstruction \cite{richardson20163d,richardson2017learning,sela2017unrestricted}.

In particular, we constructed a dataset of 3D facial human scans.
by aligning the facial geometries we are able to map the facial textures into 2D images using a universal parametrization. These images form the training data for a GAN of facial textures which is used to produce new plausible textures. Finally, each texture is coupled with a tailored geometry by learning the relation between texture and geometry in the dataset. To the best of our knowledge, the suggested model is the first to realistically synthesize both texture and geometry of human faces.
Although in this paper we apply our methodology to human faces, the general framework is not limited to this problem alone.
%

The rest of the paper is arranged as follows.
In \autoref{sec:3dmm} we describe the 3D morphable model (3DMM) which we use throughout the paper.
In \autoref{sec:pipeline} we describe our main data processing pipeline.
In \autoref{sec:gan} we describe the generative adversarial networks we used in the proposed pipeline.
In \autoref{sec:geometry_gen} we describe several methods of generating plausible geometry for a given texture.
In \autoref{sec:expermintal} we describe our experimental evaluations of our model.
In \autoref{sec:discussion} we review the main paper contributions as well as discuss our experimental results and their conclusions.

\section{3D Morphable Model}
\label{sec:3dmm}
In \cite{blanz1999morphable} Vetter and Blanz introduced a model by which the geometric structure and the texture of human faces are linearly approximated as a combination of principal vectors.
This linear model, known as the {\em 3D Morphable Model} (3DMM), was constructed by 3D scanning of several hundreds of subjects and computing dense registration between them.
Classical principal component analysis was applied to the
corresponding scans in order to obtain the principal vectors.
Then, in order to estimate the 3D face given its 2D projection, they proposed to use an analysis-by-synthesis approach, which alternates between rendering the projection and re-estimating the 3D geometry, texture, and illumination parameters in a gradient descent optimization scheme.

\begin{sloppypar}
In the 3DMM model, a face is represented by a geometry vector \mbox{$g = (\hat x^1, \hat y^1, \hat z^1, \hat x^2,...\hat y^m, \hat z^m) \in \mathbb{R}^{3m}$}  and a texture vector \mbox{$t = (\hat r^1, \hat g^1, \hat b^1, \hat r^2,...\hat g^m, \hat b^m) \in \mathbb{R}^{3m}$} that contain the coordinates and colors of its $m$ vertices, respectively.
Given a set of $n$ faces, each represented by geometry $g_i$ and texture $t_i$ vectors, construct the $3m \times n$ matrices $G$ and $T$ by grouping all geometry and texture vectors into their columns, in the same order.
Since all faces are in correspondence, Principal Component Analysis (PCA) \cite{jolliffe1986principal} can be applied in order to model the data in a basis representation.
To that end, denote by $V_g$ and $V_t$ the $3m \times n$ matrices that contain the left singular vectors of $\Delta G = G - \mu_g\mathbbm{1}^T$ and $\Delta T = T - \mu_t\mathbbm{1}^T$, respectively, where $\mu_g$ and $\mu_t$ are the average geometry and texture of the faces and $\mathbbm{1}$ is a vector of ones.
\end{sloppypar}

We assume that the columns of $V_g$ and $V_t$ are ordered by the value of the singular values in a descending manner.
The texture and geometry of each face in the model can then be defined by the linear combination
\begin{equation}
g_i = \mu_g + V_g\alpha_{g_i},\ \ \  t_i = \mu_t + V_t\alpha_{t_i},
\end{equation}
where $\alpha_{g_i}$ and $\alpha_{t_i}$ are the coefficients vectors,  obtained by $\alpha_{g_i} = V_g^T(g_i - \mu_g)$ and $\alpha_{t_i} = V_t^T(t_i - \mu_t)$.
Following this formulation, one can generate new faces in the model by changing the texture and geometry coefficients.
In order to obtain plausible faces from the model, the distribution of faces is assumed to follow a multivariate normal distribution, so that the probability for a coefficient vector $\alpha$ is given by
\begin{equation}
P(\alpha) \sim \exp\left \{-\frac{1}{2}\alpha^T\Sigma^{-1}\alpha\right \},
\label{eq:3dmm_dist}
\end{equation}
where $\Sigma$ is a covariance matrix that can be empirically estimated from the data, and for simplicity assumed to be diagonal.

Lastly, in order to obtain robust results when synthesizing new faces, only the first $k \ll n$ basis vectors and corresponding coefficients should be considered in the linear combination. That is, higher order basis vectors, corresponding to smaller PCA singular values, do not have enough data to faithfully estimate their values.
Moreover, the number of $3D$ faces in a given dataset is limited and typically cannot cover all high resolution geometries and textures.
Taken together, the above linear combination would most likely result in a smooth geometry and texture.

\begin{figure}
\centering
\begin{subfigure}[t]{0.49\linewidth}
\includegraphics[width=1\linewidth]{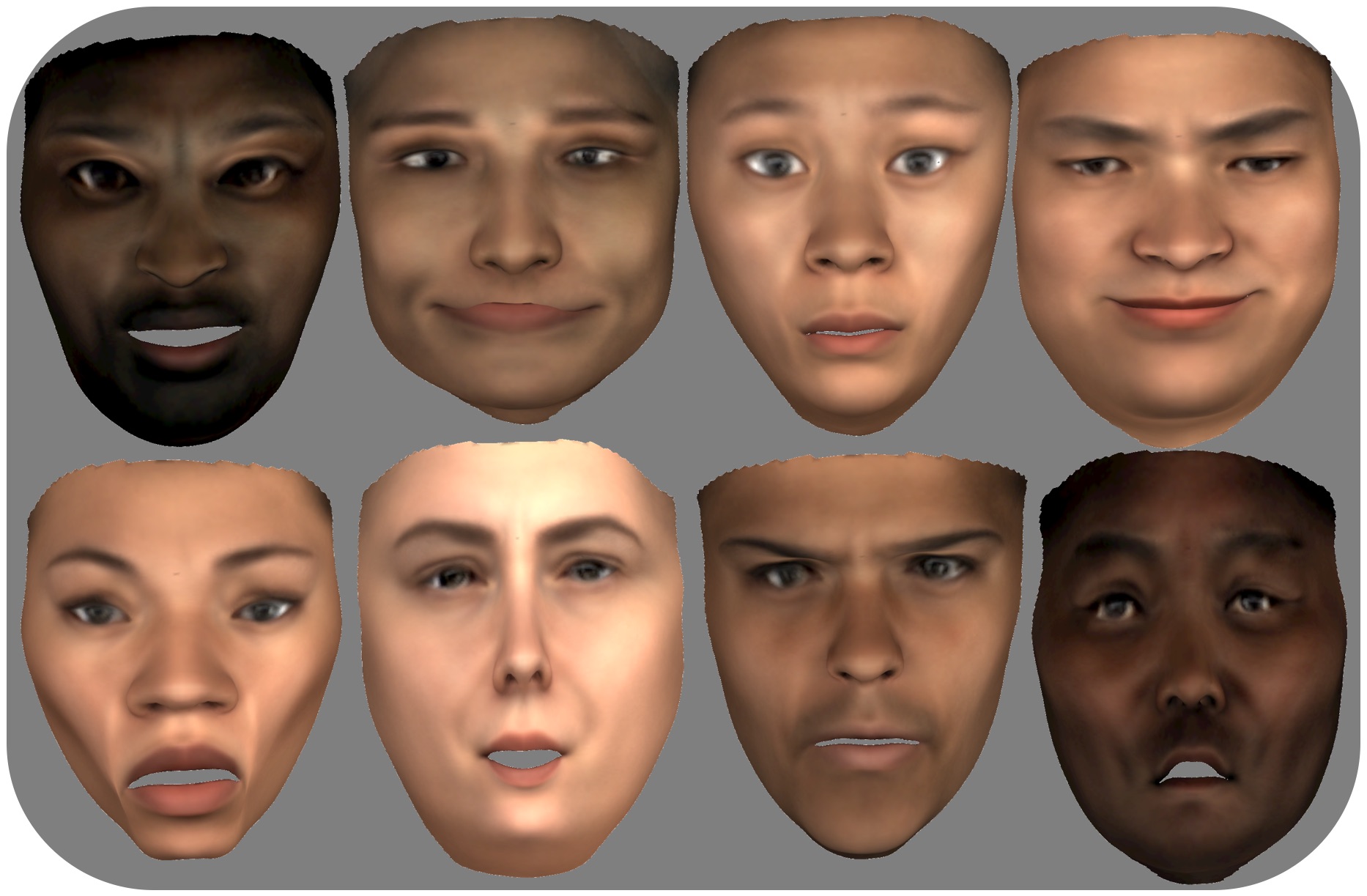}
\caption{}
\label{fig:3dmm_results}
\end{subfigure}
\begin{subfigure}[t]{0.49\linewidth}
\includegraphics[width=0.49\linewidth,trim={2cm -3cm 2cm 2cm}, clip]{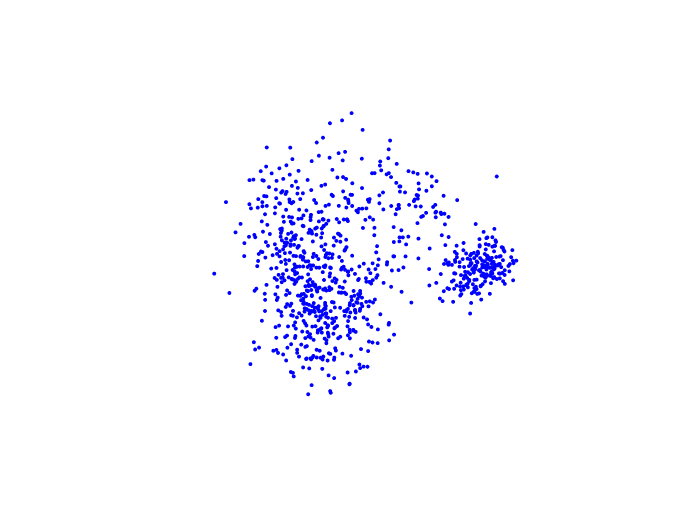}
\includegraphics[width=0.49\linewidth,trim={2cm -3cm 2cm 2cm}, clip]{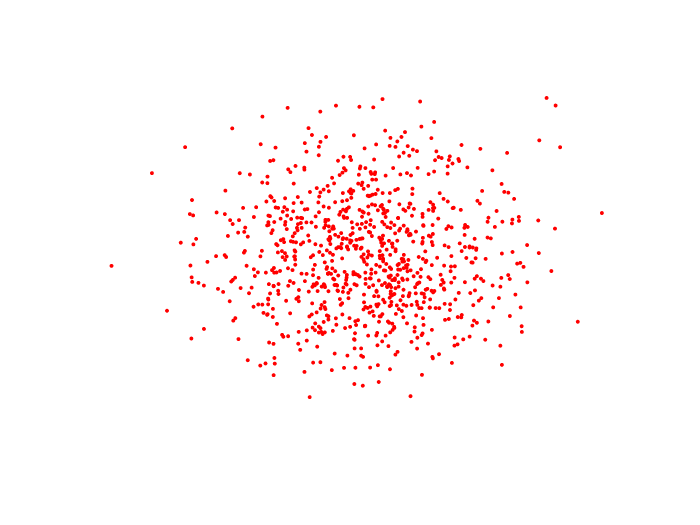}
\caption{}
\label{fig:face_distributions}
\end{subfigure}
\caption{(a) Faces synthesized using the 3DMM linear model. (b) First two PCA coefficients of real and 3DMM generated faces. Left - Real faces distribution. Right - 3DMM faces distribution.}
\end{figure}

Recently, the 3DMM model was integrated with convolutional neural networks for data augmentation and for recovering pose variations of faces in images \cite{sela2017unrestricted,richardson20163d,richardson2017learning,gecer2018semi}.
However, faces rendered using the above PCA model tend to be smooth and non-realistic.
Using them for data augmentation would require additional steps such as  transfer learning or designing additional networks to bridge this gap.
Additionally, the multivariate normal distribution assumption rarely follows the true distribution of the data.
In \autoref{fig:3dmm_results} we show examples of faces synthesized using the 3DMM model while considering $k=200$ basis vectors with corresponding random coefficients for each face.
In \autoref{fig:face_distributions}, we plot the first two coefficients of real faces, computed by projecting the faces onto the 3DMM basis, and compare them to the coefficients of the synthesized 3DMM faces, showing the gap between the real and the synthesized distributions.

\section{Data Acquisition Pipeline}
\label{sec:pipeline}
\begin{figure}
\centering
\includegraphics[width=0.22\linewidth,trim={1cm 1cm 0cm 1cm}, clip]{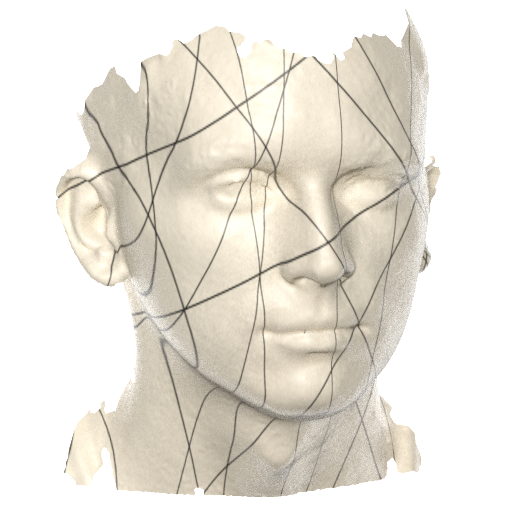}
\includegraphics[width=0.22\linewidth,trim={0cm 1cm 1cm 1cm}, clip]{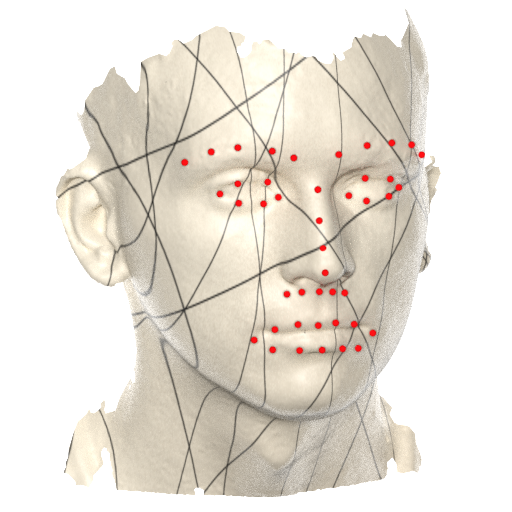}
\includegraphics[width=0.22\linewidth,trim={2cm 1cm 0cm 1cm}, clip]{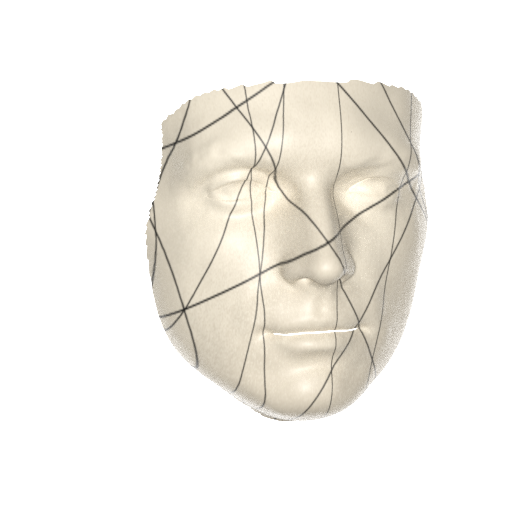}
\includegraphics[width=0.22\linewidth,trim={0cm 1cm 1cm 1cm}]{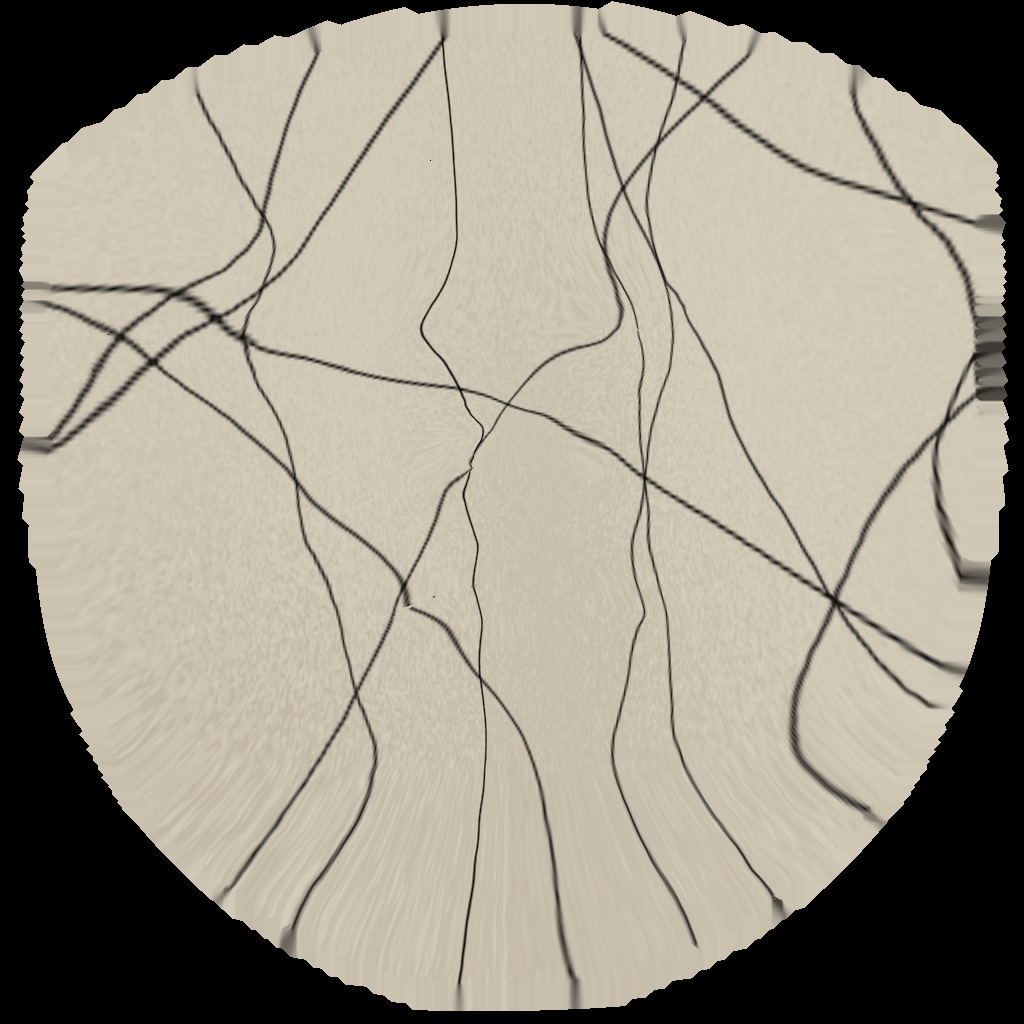}\\
\caption{Data preparation pipeline, left to right: Real scan geometry with demo texture. 3D facial landmarks are added. A template mesh is deformed to fit the scan guided by the landmarks. Texture is transferred onto the deformed template. The texture is mapped to a 2D plane using a universal mapping for the entire dataset.}
\label{fig:pipeline}
\end{figure}

Our main data acquisition pipeline was designed to align 3D scans of human faces vertex to vertex, and map their textures onto a 2D plane using a predefined universal transformation.
This process is comprised of four main stages, depicted in \autoref{fig:pipeline}:
data acquisition, landmark annotation, mesh alignment and texture transfer.
In the following section we describe each stage in detail.


Our data construction pipeline begins with the acquisition of high resolution geometric facial scans of human subjects.
Due to privacy concerns, we are not permitted to share or display the raw data directly. 
Motivated by \cite{booth2018large,booth20163d}, we collected roughly $5000$ scans from a wide variety of ethnic, gender, and age groups, using a 3DMD\texttrademark scanner.
Each subject was asked to perform five distinct expressions including a neutral one.
The acquired data went through a manual selection process which is intended to filter out corrupted  meshes. 

Once the data was collected, we proceeded to produce $3D$ facial landmark annotations in order to guide the subsequent alignment stage.
We used Multi-pie $68$ standard facial feature points \cite{gross2010multi}, out of which we discarded the jaw and inner lip due to their instability.
The remaining $43$ landmarks were added to the meshes semi-automatically by rendering the face and using a pre-trained facial landmark detector \cite{menpo14}\cite{dlib09} on the 2D images. The resulting 2D landmarks are back-projected onto the 3D mesh.
In order to achieve an even more reliable annotation process, a human annotator manually corrected the erroneous landmarks.

During the alignment stage of the pipeline we performed a vertex to vertex correspondence between each scan and a facial template mesh.
The alignment was conducted by deforming a template face mesh according to the geometric structure of each scan, guided by the previously obtained facial landmark points.
The deformation process minimizes the energy term in  \cite{blanz1999morphable}.
The energy is made up of $3$ terms which contribute to the final alignment.
The first term accumulates the distances between the facial landmark points on the target and on the template.
The second term aims to minimize the distance between all the mesh points on the template to the target surface.
The third term serves as a regularization, and penalizes non-smooth deformations.
The energy is minimized using gradient descent until convergence.
This alignment process is the cornerstone of our data preparation pipeline.

\begin{figure}
\centering
\includegraphics[width=0.18\linewidth]{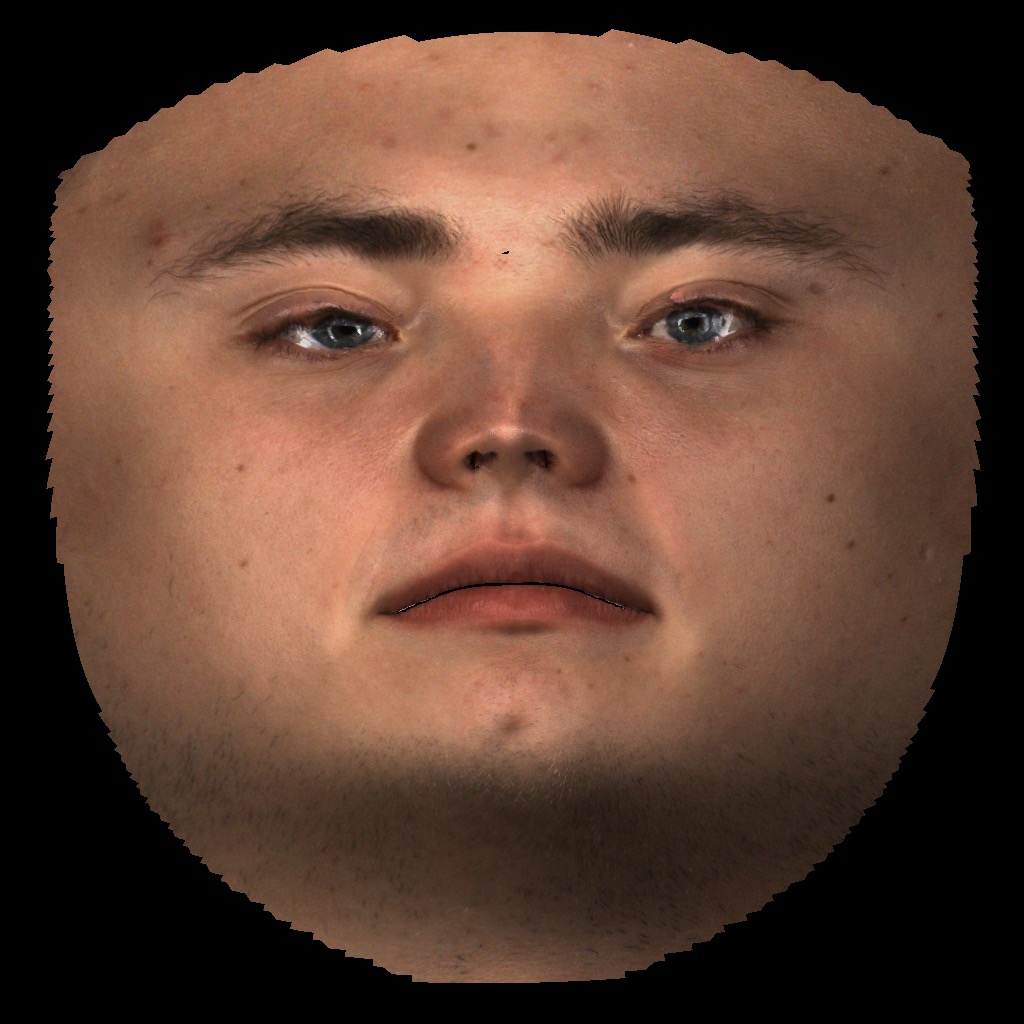}
\includegraphics[width=0.18\linewidth]{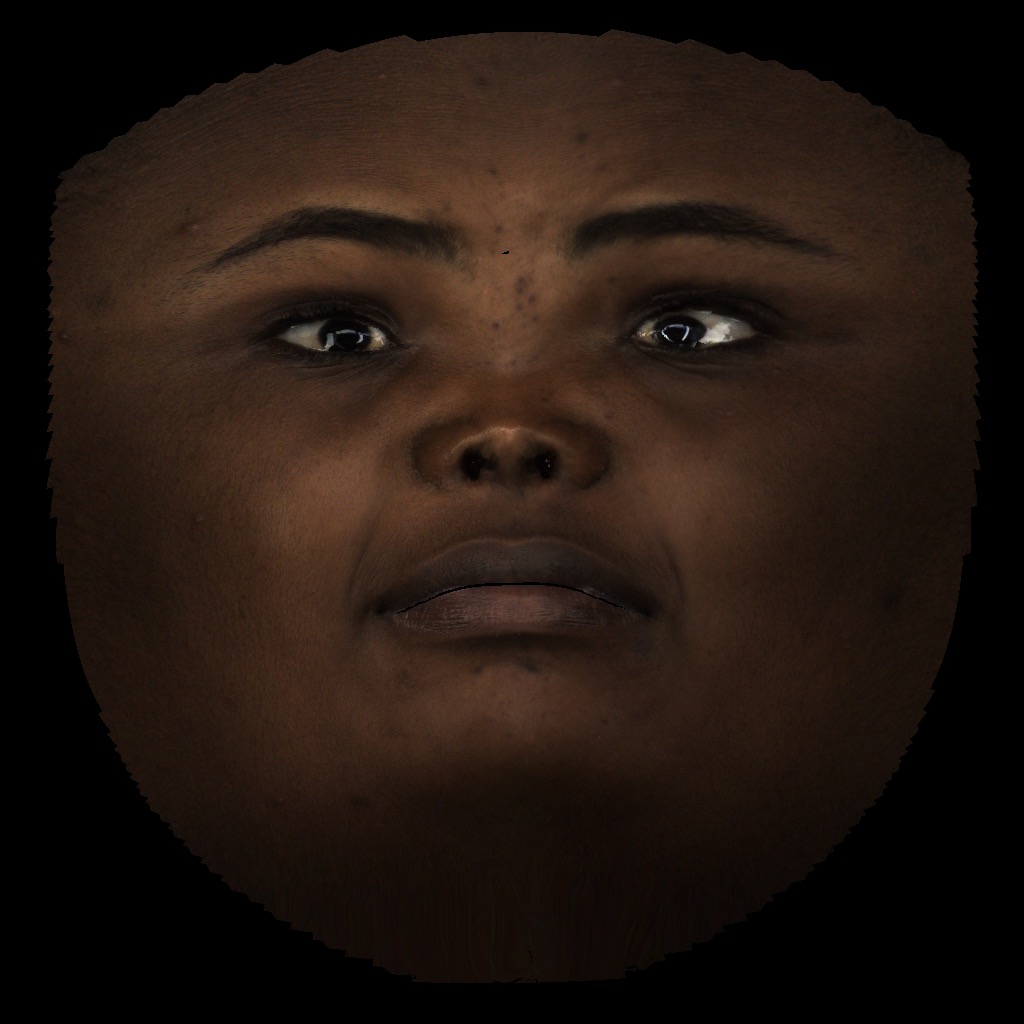}
\includegraphics[width=0.18\linewidth]{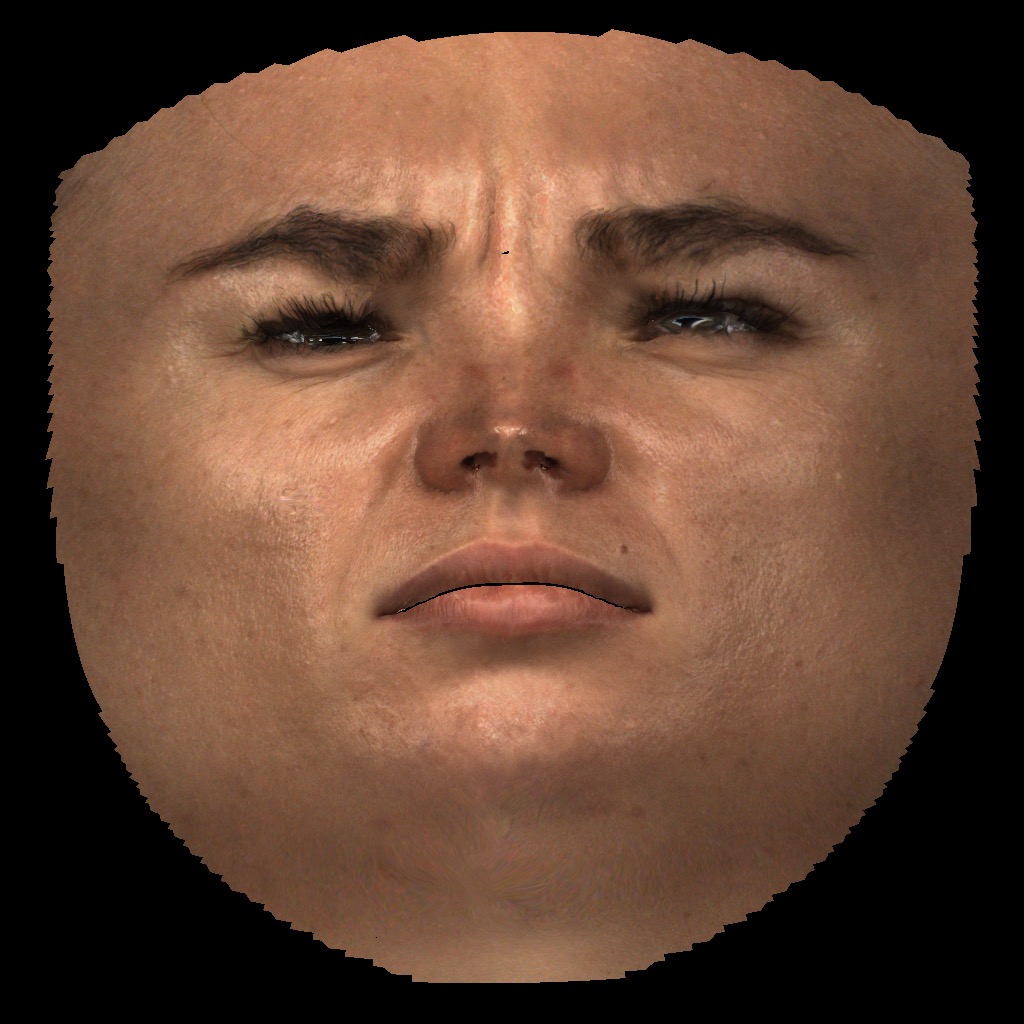}
\includegraphics[width=0.18\linewidth]{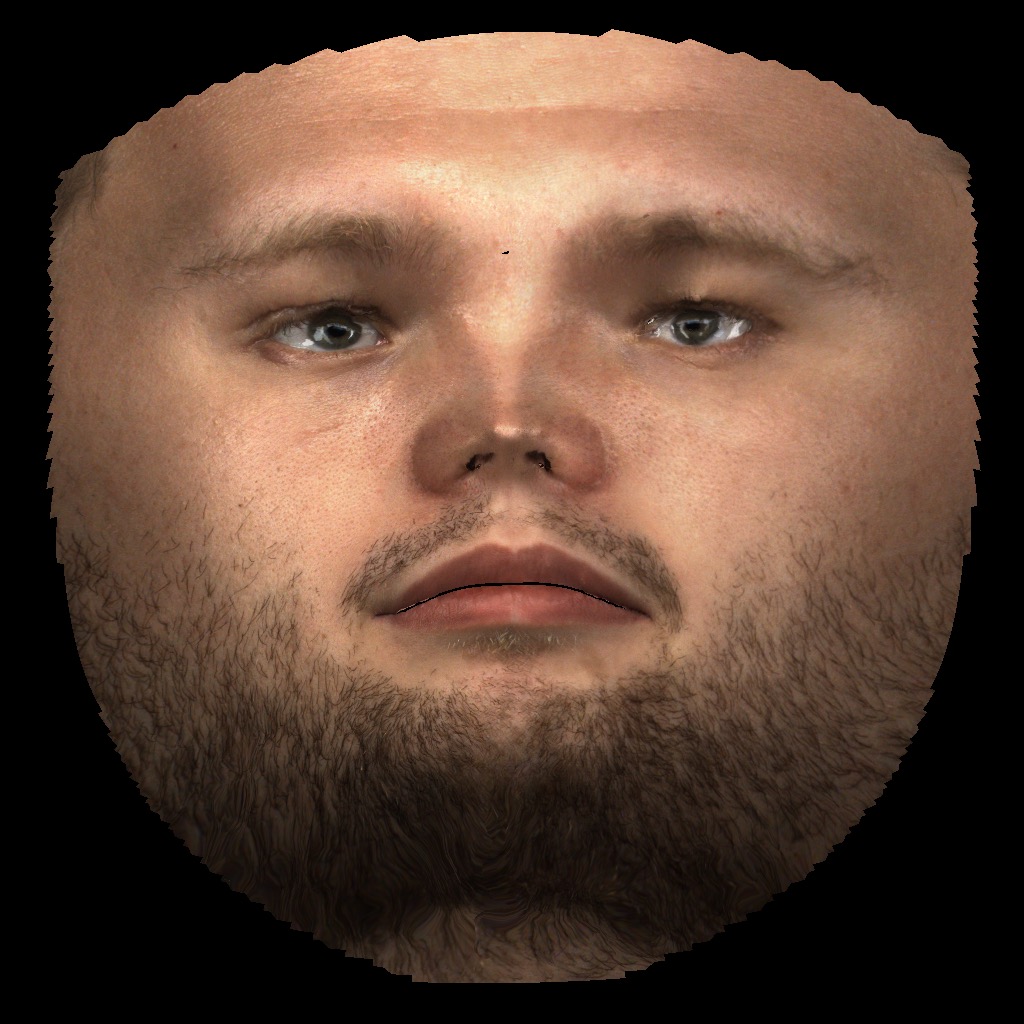}
\caption{Flattened aligned facial textures.}
\label{fig:real_textures}
\end{figure}

Once the deformed template is properly aligned with the original scan, the texture is transferred from the scan to the template using a ray casting technique built into the animation rendering toolbox of Blender \cite{blender}. The texture is then mapped from the template to a 2D plane using a predefined universal mapping that was constructed once.
As a result, the textures of all scans are semantically aligned under a universal mapping.
The semantic alignment simplifies the network learning process, since the data is invariant with respect to the locations of the facial parts within the image.
In Figure \ref{fig:real_textures} we show the resulting mapped textures of the dataset.

\section{GAN}
\label{sec:gan}
Generative models which are able to mimic examples from a high dimensional data distribution are recently gaining popularity. Such models are tasked with producing novel examples of data from a learned distribution. The first to propose such a model which is based on a deep neural network was \cite{goodfellow2014generative}, who dubbed the term generative adversarial network (GAN). Recent advances in GANs have shown promise in synthesis of audio signals \cite{oord2016wavenet}, images \cite{karras2017progressive}, and image to image translation \cite{isola2017image,zhu2017unpaired}.
Here, we use a GAN in a novel way to synthesize high resolution realistic facial textures which can be fitted to 3D facial models. In this section we will describe briefly the main idea and architecture of GANs in general, and more specifically that of \cite{karras2017progressive} which we adopt for our purpose.

A GAN is a special form of convolutional neural network which is designed to generate data samples which are indistinguishable from the training set. The GAN is comprised of two separate networks which are competing against each other. The generator network aims to produce novel examples while the discriminator network aims to distinguish between the generated and real examples.
The generator network takes as input a random high dimensional normalized latent code and produces a sample of the same dimension as the data which makes up our training set.
Ideally, the implemented loss should penalize deviation from the true data distribution and encourage generated examples which follow it.
This loss, however, is highly complex and impractical to design manually.
The way to circumvent this problem is to construct the loss function as a dynamic network which continually improves its assessment on how to distinguish between the fake and real data samples.
This is why the discriminator network which is used as part of the loss is trained alongside with the generator.
The key of the GAN method is that the discriminator produces a gradient which can be used to update the generator weights so that the generated samples are better at confusing the discriminator.
This results in a race between the generator and discriminator, constantly improving each other as the training progresses.
The typical loss of a GAN can be formulated as the min max loss
%
\begin{equation}
\footnotesize
\label{eq:minimaxgame-definition}
\min_G \max_D V(D, G) = \mathbb{E}_{x \sim p_{\text{data}}(x)}[\log D(x)] +
\mathbb{E}_{z \sim p_{z}(z)}[\log (1 - D(G(z)))],
\end{equation}
where $G$ and $D$ denote the generator and discriminator, $x$ denotes the true data, and $z$ denotes the latent space representation. More sophisticated loss functions have recently shown success in the training process. Some noteworthy examples are Wasserstein loss \cite{gulrajani2017improved,arjovsky2017wasserstein} and least squares loss \cite{mao2017least}, which apply different metrics to the computation of distances between data distributions.

\begin{figure}
\centering
\includegraphics[width=0.18\linewidth]{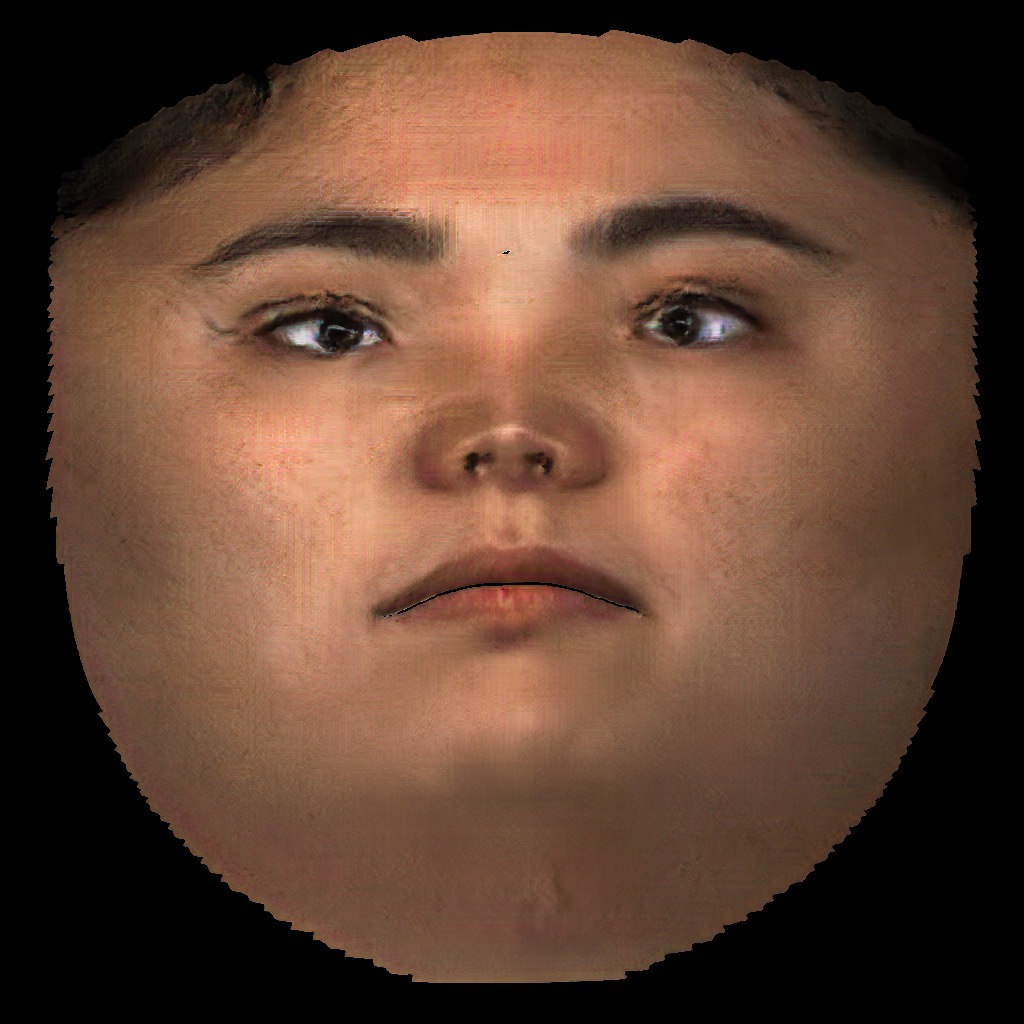}
\includegraphics[width=0.18\linewidth]{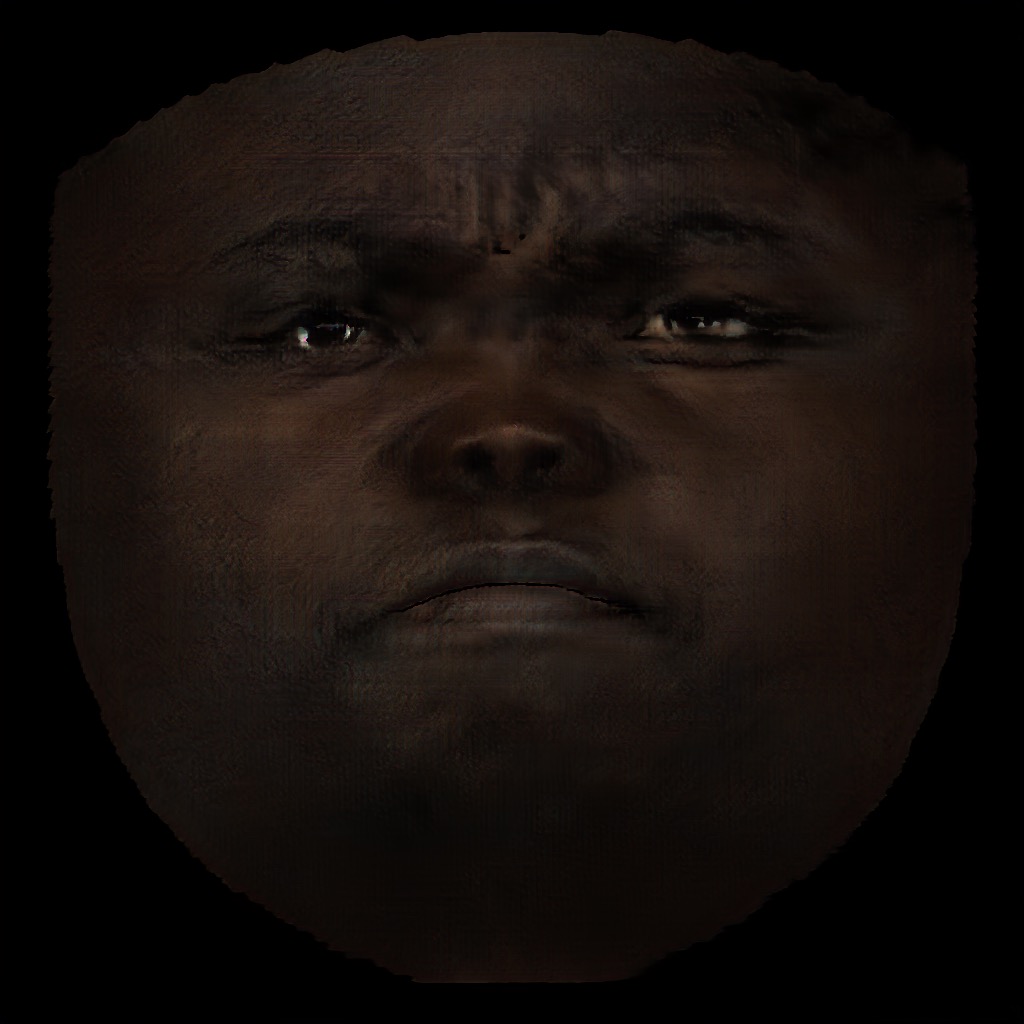}
\includegraphics[width=0.18\linewidth]{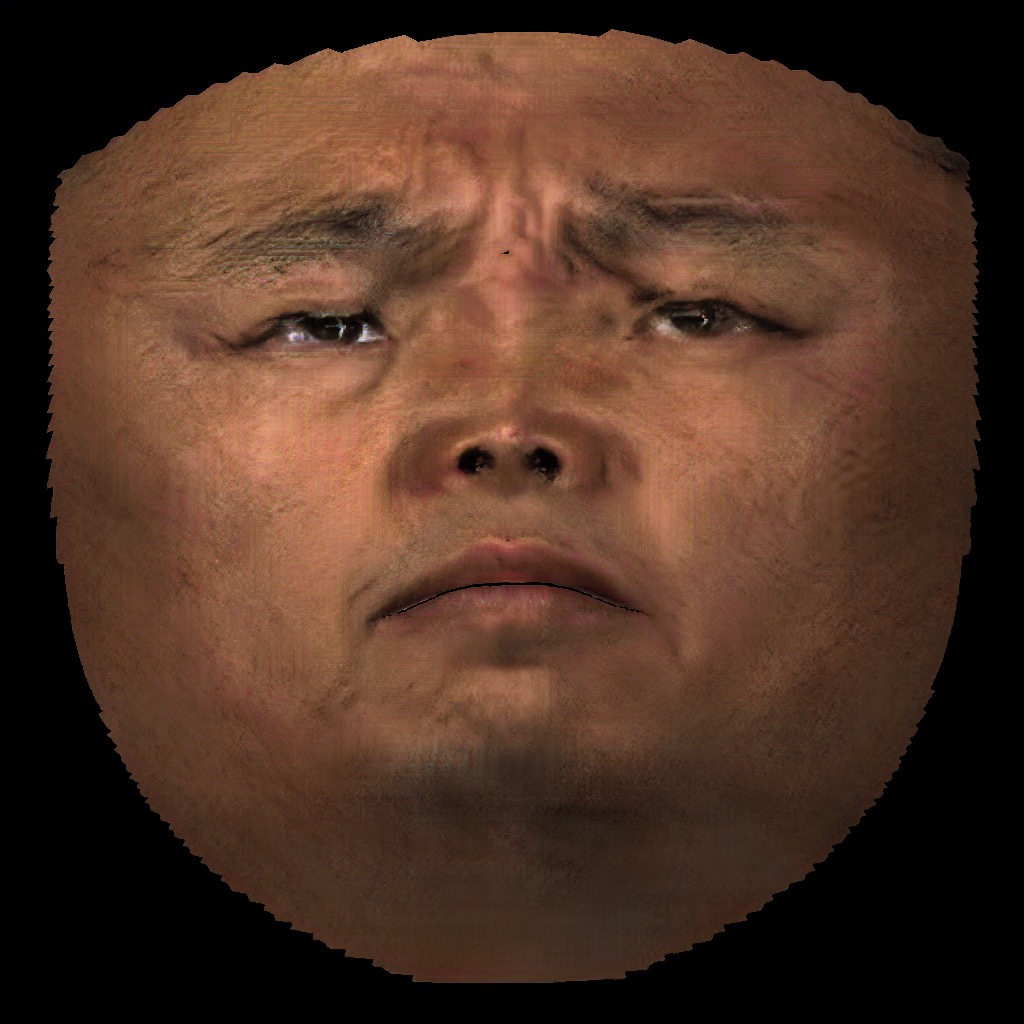}
\includegraphics[width=0.18\linewidth]{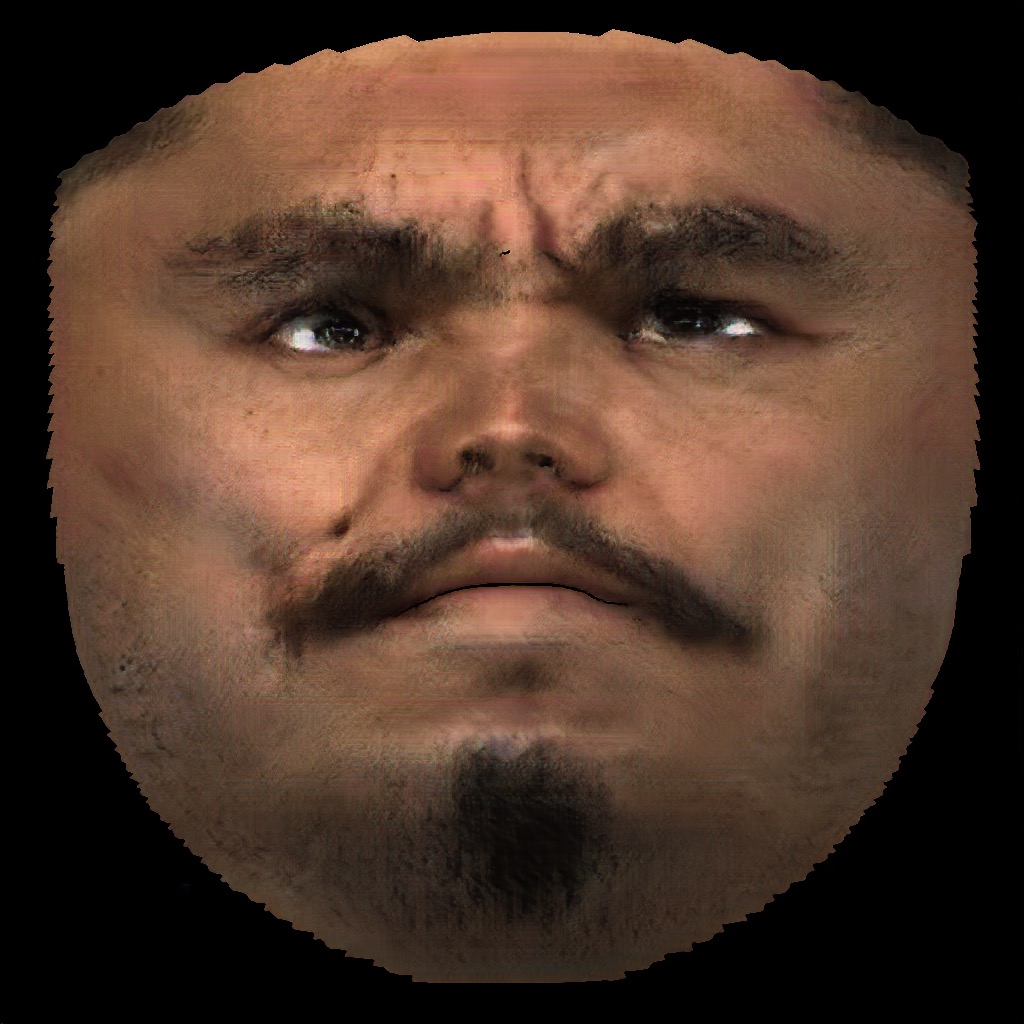}
\caption{Facial textures synthesized by GAN.}
\label{fig:GAN_results}
\end{figure}

Here we use a successful implementation of GAN dubbed progressive growing GAN \cite{karras2017progressive}.
This architecture combines several novel contributions which improve the training stability and the resulting image quality.
The core idea is to construct the generator and discriminator as symmetric networks.
The generator progressively increases the resolution of the feature maps at each stage until reaching the output image size, while the discriminator gradually reduces the size back to a single output.
The training starts from the lowest resolution feature maps, and is guided by low resolution versions of the input data.
After a stabilization period, a new layer is added by mixing an up-scaled version of the output emanating from the lower level feature maps, with the higher level output.
The mixing coefficient gradually gives more importance to the higher level features at the output, until the contribution of the up-scaled lower layer is discarded completely.
At this stage the new layer goes through a stabilization phase, and so on.
According to \cite{karras2017progressive}, this is the main contribution to the training stabilization and improvement of results.
Here, we trained the aforementioned GAN to learn and imitate the previously obtained aligned facial textures. The new synthetic facial textures generated by the GAN are shown in \autoref{fig:GAN_results}.


\section{Synthesizing the Geometry}
\label{sec:geometry_gen}
\begin{figure}
\centering
\begin{subfigure}[c]{0.37\linewidth}
\centering
\includegraphics[width=1\linewidth,trim={0cm 0cm 0cm -6cm}, clip]{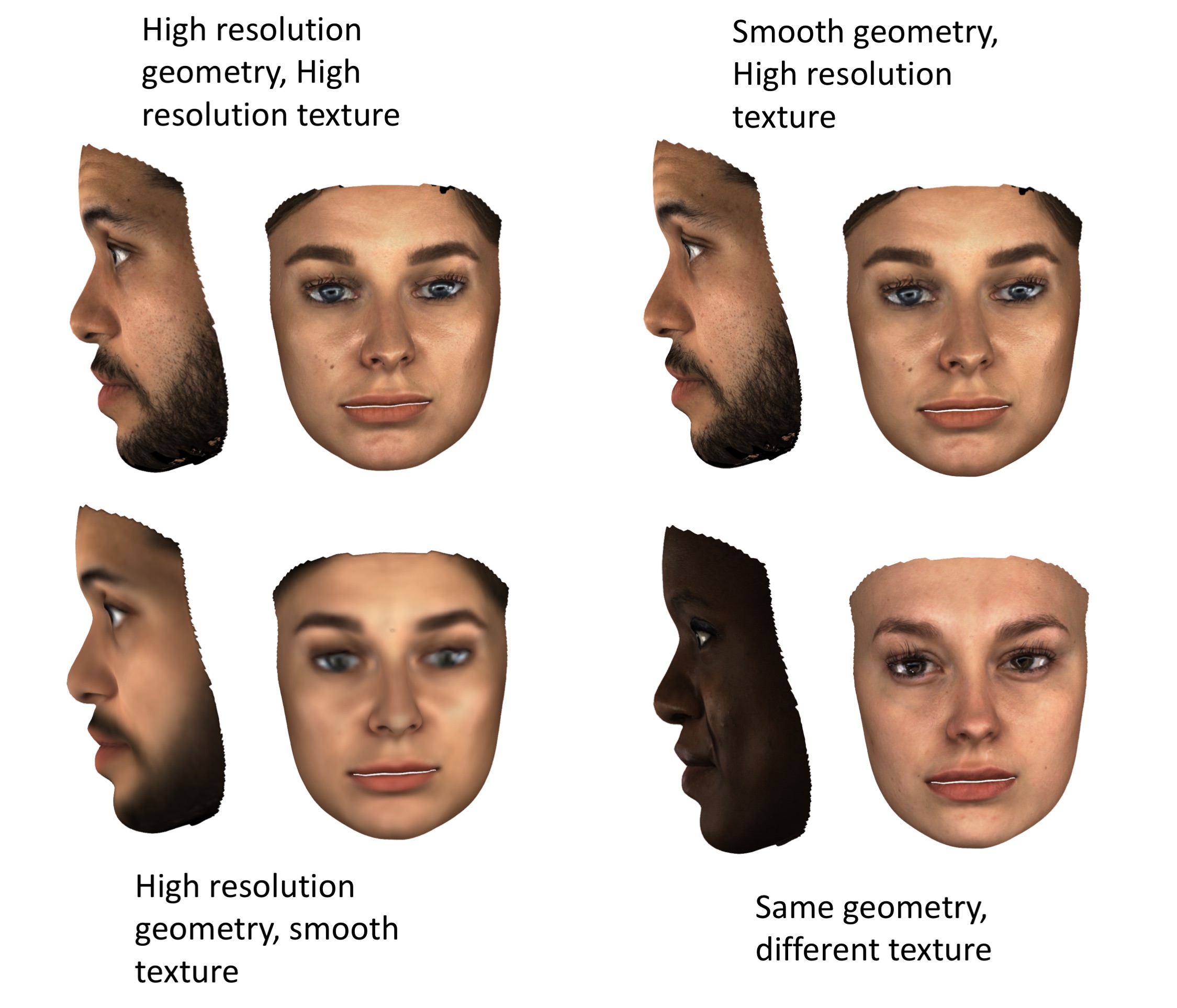}
\caption{}
\label{fig:texture_geometry_compare}
\end{subfigure}
\hspace{1cm}
\begin{subfigure}[c]{0.45\linewidth}
\centering
\includegraphics[width=1\linewidth,trim={0cm -2.5cm 0cm -0.5cm}, clip]{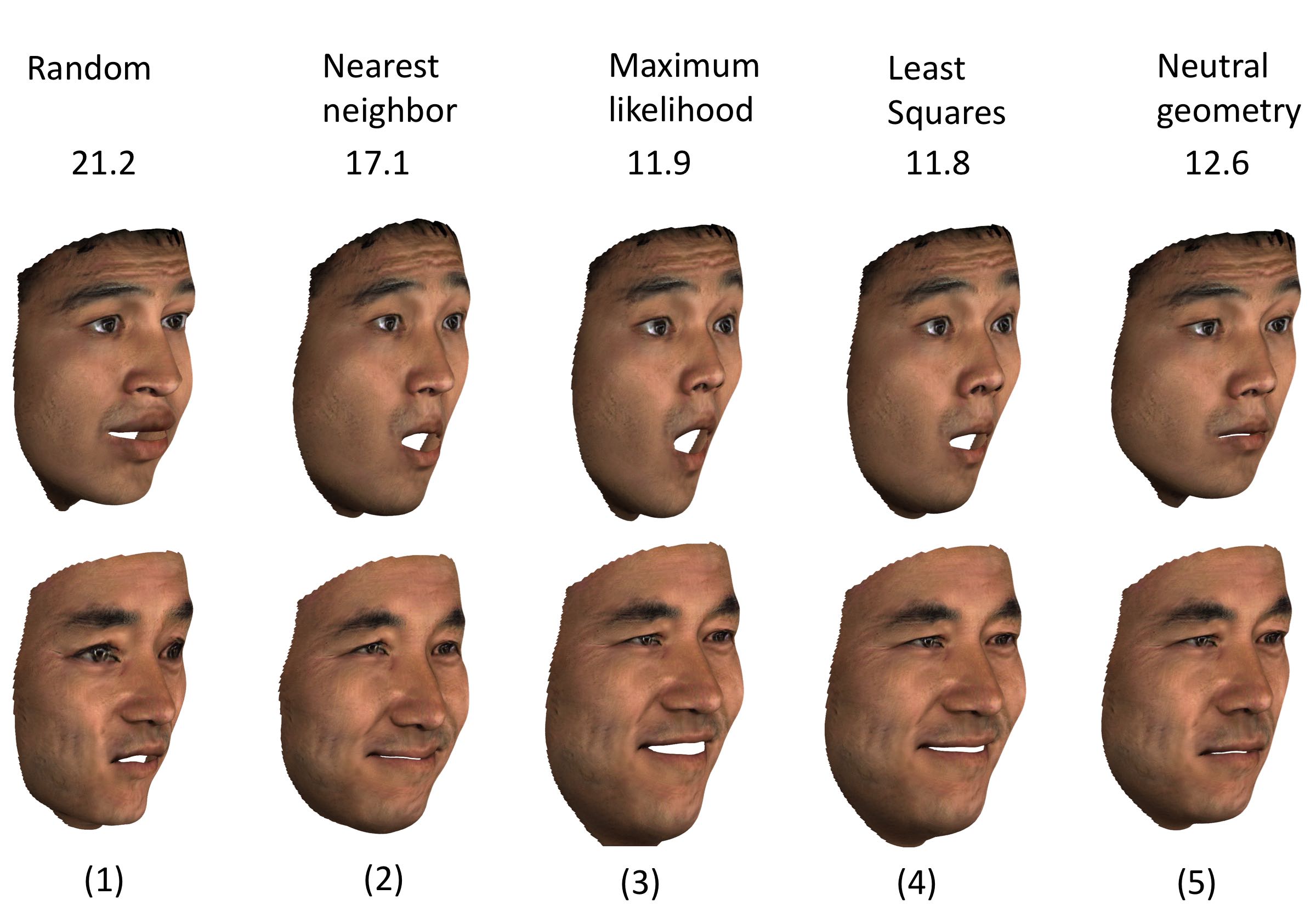}
\caption{}
\label{fig:geometry_synthesis}
\end{subfigure}
\caption{\footnotesize
 Left: Perceptual quality comparison between reduction in geometry detail (\mbox{$k = 200$} geometry basis vectors) vs reduction in texture detail. Geometric detail loss is very difficult to perceive while texture detail loss is detrimental to the final outcome.
 Right: Two synthesized textures mapped onto different geometries.
 Each geometry is produced by a method discussed in \autoref{sec:geometry_gen}. (1) random geometry, (2) real geometry from nearest real texture, (3) ML geometry, (4) LS geometry, (5) Neutral geometry. The neutral geometry was obtained using the LS method. The $L_2$ geometric error tested by 10-fold cross validation is presented for each method at the top.
 }
\end{figure}

A popular way to deal with 3D textured object representations is to consider a discrete version of their geometry, such as a polygon mesh, while keeping their texture resolution high by mapping the texture from an image to the mesh.
The observation that the geometry discretization of the object has a small impact on its appearance can be exploited for the sake of our geometry synthesizer.
\autoref{fig:texture_geometry_compare} demonstrates that objects with smooth geometry and high resolution texture appear to be visually similar to their high resolution geometry and texture versions.
This shows that the facial texture has a greater influence on perceived appearance than their geometry.
Following this assumption, we propose to exploit the 3DMM discussed in \autoref{sec:3dmm} to generate the geometries of our faces as a linear combination of the first $k \ll n$ geometry basis vectors as
\begin{equation}
    g = \mu_g + \sum_{i=1}^{k}\alpha_{gi}v_{gi},
\end{equation}
where $v_{gi}$ is the $i$-th vector of the geometry basis $V_g$ and   $\{\alpha_{gi}\}_{i=1}^k$ are the coefficients that define the geometry.

In this section we explore several possible methods to find plausible geometry coefficients for a given texture. In each subsection we present one of the methods and briefly discuss their strengths and weaknesses. \autoref{fig:geometry_synthesis} makes a qualitative and quantitative comparison between the various methods.
For this comparison, we measure the distance between recovered geometries and the true corresponding geometries on a held-out set of real samples that were not included in the geometry recovery process, using a 10-fold cross validation. The  distance is measured by the average of $\|g_r - g_t\|_{L_2}$
for all faces, where $g_r$ and $g_t$ are the recovered and true geometries.
The outcome of the comparison is presented for each one of the methods in \autoref{fig:geometry_synthesis}.

\subsection{Random Geometries}
The simplest way to choose the geometry coefficients is by exploiting the multivariate normal distribution assumption of the 3DMM in order to pick random coefficients.
Following the formulation in \autoref{eq:3dmm_dist}, the probability of the coefficient $\alpha_i$ is
\begin{equation}
P(\alpha_i) \sim \exp\left \{-\frac{\alpha_i^2}{2\sigma_i^2}\right \},
\end{equation}
where $\sigma_i^2$ is the $i$-th eigenvalue of the covariance matrix of $\Delta G$, which can be also computed more efficiently as $\sigma_i^2 = \frac{1}{n}\delta_i^2$, where $\delta_i$ is the $i$-th singular value of $\Delta G$.
Following that, we compute the singular values $\{\delta_i\}_1^k$ of $\Delta G$ and randomize a vector of coefficients from the above probability.
The problem with picking a random geometry for each face is that the correlation between the texture and the geometry is ignored.
For instance, facial texture could indicate a specific ethnicity or gender which are related to specific geometric traits.
The advantage of this method is that it is very simple and fast, and many identities can be created out of a single generated texture.

\subsection{Nearest neighbors}

Given a facial texture, a simple way to obtain a geometry that is both plausible and likely to fit the texture is by finding a face in the data with a similar texture and then taking its geometry.
Here, given a new facial texture generated by the proposed pipeline, we find the face in the data with the most similar texture, in terms of $L_2$ norm between the 3DMM texture coefficient, and use its geometry for our synthetic face. This would only require storing the 3DMM coefficients of the training data.
The resulting geometry would most likely fit the texture and, moreover, will not loose its high frequencies.
Nevertheless, geometries generated in this manner would be constrained to a small set of possible geometries.
Additionally, the geometry will indeed retain accurate geometric details, but these will not coincide with the subtle details of the generated texture.
For example, a texture of mole in the face would not have a corresponding curved geometry.

\subsection{Maximum likelihood approach}
Returning to the 3DMM formulation, one suitable way to obtain geometry coefficients that most likely fit the texture is by using a maximum likelihood approximation.
The mathematical formulation regarding this approach is detailed in the supplementary material.
\subsection{Least squares approach}
\label{sec:LS}
The maximum likelihood approach is typically used when a small amount of data is available and one can have some assumptions on the distribution of the data. When sufficient data samples are available, it is usually more beneficial and straightforward to learn or estimate parameters using a least squares minimization scheme.
We start from the original 3DMM model defined in \autoref{sec:3dmm}.
Given a texture coefficient vector $\alpha_t$, we would like to estimate a plausible geometry coefficient vector $\alpha_g$.
To that end, we group all coefficient vectors $\alpha_t$ and $\alpha_g$ from our data into the columns of the matrices $A_g$ and $A_t$ and search for a matrix $W$ such that
\begin{equation}
    \mbox{loss}(W) = \|W^TA_t - A_g\|_{F}
\end{equation}
is minimized.
The vanishing gradient of the above least squares minimization problem,
 yields the solution $W$ given as a closed form by
\begin{equation}
    W^* = (A_tA_t^T)^{-1}A_tA_g^T = A_t^{+}A_g^T.
\end{equation}

Given a texture $t$ of a new synthesized face, one can first compute the texture coefficient vector $\alpha_t$ as $\alpha_t = V_t^T(t - \mu_t)$, then compute its geometry coefficient vector $\alpha_g$ as $\alpha_g=W\alpha_t$, and finally compute the geometry as $g = \tilde V_g \tilde \alpha_g + \mu_g$,
where $\tilde V_g$ and $\tilde \alpha_g$ hold the first $k$ vectors and coefficients of $V_g$ and $\alpha_g$.
It can be seen in \autoref{fig:geometry_synthesis} that the LS approach produces the lowest distortion results among our tested methods. For this reason and due to its simplicity, we chose to apply it during all of our subsequent experiments. It is worth mentioning that other applications may benefit from using one of the other methods according to their objective.
Our experimental results give another verification to the validity of this approach, in terms of identity distribution and variation as presented in \autoref{sec:expermintal}.

\subsection{Neutral face geometries}
\label{sec:neutral}
The data we have worked with contains faces with a neutral and four other expressions.
When estimating the geometry using each of the above methods, we consider all expressions as part of the model.
Nevertheless, in some cases one would like to only obtain faces with a neutral expression geometries.
One example for the necessity of the neutral pose face model is when the expression is modified using the Blend Shapes \cite{chu20143d} (See the Blend Shapes experiment in \autoref{sec:expermintal}). The Blend Shapes model takes as input a neutral face and adds linear combinations of facial expressions in order to span the space of possible expressions.


To estimate the geometry for each of the above methods while constraining it to a neutral expression, we suggest to simply replace any geometry $g_i$ in $G$ by the neutral geometry of the same identity in the dataset.
Then, repeat the process of any of the above methods.
In this manner, we tie each of the textures, regardless of its expression, to the neutral geometry of their identities rather than to the actual geometry which includes the non-neutral expression.

\section{Experimental results}
\label{sec:expermintal}
Throughout this section, we use the proposed texture generation model and the Least squares approach described in \autoref{sec:LS} for generating the corresponding geometries.
%
%
The main advantage of our proposed method is that it can be used to create many new identities, and each one of them can be rendered under varying pose, expression and lighting.
Given a facial texture synthesized by our system, we extracted its neutral geometry using the method described in \autoref{sec:neutral}.
We then used the Blend Shapes model as suggested in \cite{chu20143d} to add different expressions to the facial geometry.
We changed the pose and lighting and rendered 2D images to obtain numerous examples of the same identity.
The resulting images are shown in the supplementary material in figures 1-3.

The sliced Wasserstein distance (SWD) is an efficiently computable randomized approximation to earth-movers distance which can be used to measure statistical similarities between images \cite{rabin2011wasserstein}.
A small SWD indicates that the distribution of the patches is similar in both appearance and variation.
The authors in \cite{karras2017progressive} used SWD to measure the distance between the training distribution and the generated distribution of their GAN in different scales, and compared them to results produced by various competing methods.
More specifically, the SWD between the distributions of patches extracted from the lowest resolution images is indicative of similarity in large-scale image structures, while the high resolution patches encode information about pixel-level attributes.
Inspired by this notion, we used the SWD to measure distances between distributions of our training and generated images in different scales.
The results of this experiment are shown in \autoref{tbl:SWD-table}.
The table demonstrates that the textures generated by our model are statistically closer to the real data than those generated by 3DMM.
\begin{table}[]
\centering
{\small
\begin{tabularx}{\linewidth}{lXXXXXXXX}
Resolution & 1024 & 512 &  256&  128&  64& 32& 16& avg\\ \hline
Real & 3.53 &2.98  &3.75 &2.6 & 2.75 & 2.5 &1.63 &2.82\\
\textbf{{Proposed}} &  20.62 & 10.02  &8.52 &7.58 &7.75  &9.89 & 21.77 & 12.31\\
PCA &326& 137& 42& 19.3& 11.74& 22.86& 72.51& 90.52\\
\end{tabularx}
}
\caption{Sliced Wasserstein distance (SWD) \cite{rabin2011wasserstein} measured over extracted patches from the real and generated textures. The columns show SWD for patches extracted at different image resolutions, and the final column shows the average SWD over all resolutions.}
\label{tbl:SWD-table}
\end{table}

\begin{figure}
\begin{center}
\includegraphics[width=0.24\linewidth]{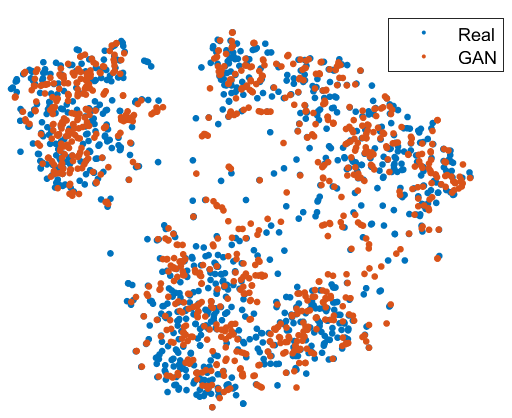}
\includegraphics[width=0.24\linewidth]{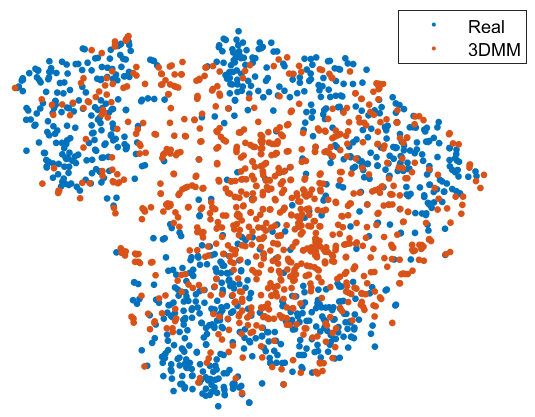}
\includegraphics[width=0.24\linewidth]{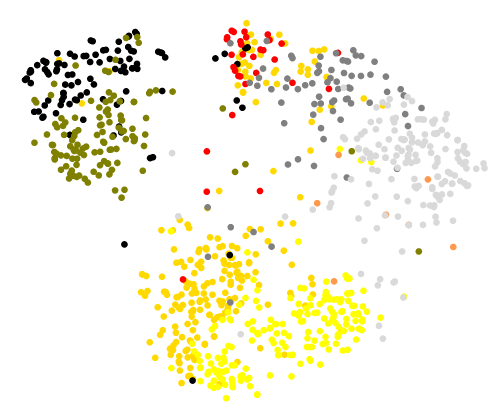}
\includegraphics[width=0.15\linewidth]{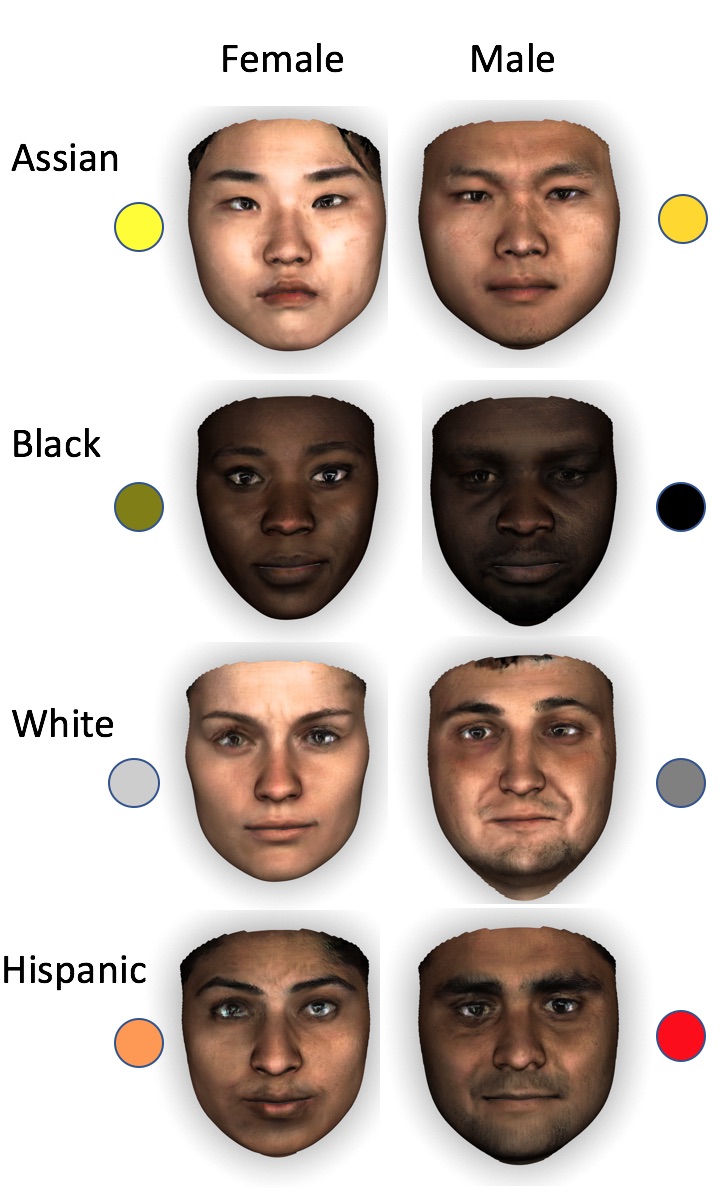}
\end{center}
\caption{From left to right:
T-SNE \cite{maaten2008visualizing} embedding of real identities versus GAN synthetic identities,
T-SNE embedding of real identities versus 3DMM identities,
clusters of real data according to race and gender, and synthetic samples conforming to each cluster.}
\label{fig:TSNE_experiment}
\end{figure}

In order to visualize and compare between the distributions of the real and generated data, we used a popular dimensionality reduction process termed T-SNE \cite{maaten2008visualizing}.
We generated textures and geometries according to our model, as well as according to the 3DMM. For 3DMM, we used $200$ eigenvectors for both texture and geometry.
We then rendered the real and generated faces and fed them into a pre-trained facial recognition network based on \cite{amos2016openface}, which provided an identity descriptor for each rendered face. We used T-SNE to visualize the distribution of obtained identity vectors.
We labeled the real data according to race and gender, and found very uniform clusters.
We assigned each one of our the faces generated by our proposed model to the closest cluster's center and produced one random example from each cluster.
The results of this process are depicted in \autoref{fig:TSNE_experiment}.
The embedding clearly shows that the distribution of identities produced by the proposed pipeline is well matched to the distribution of the real identities included in the training data, and that the pipeline is able to produce data samples from each cluster (race and gender) reliably.
On the other hand, the distribution produced by 3DMM, generated by the same training data, is a uniform Gaussian which has no natural clustering.

\begin{figure}
    \centering
    \begin{subfigure}[t]{0.35\linewidth}
    \includegraphics[width = 1\linewidth,trim={2cm 0cm 2cm 0cm}, clip]{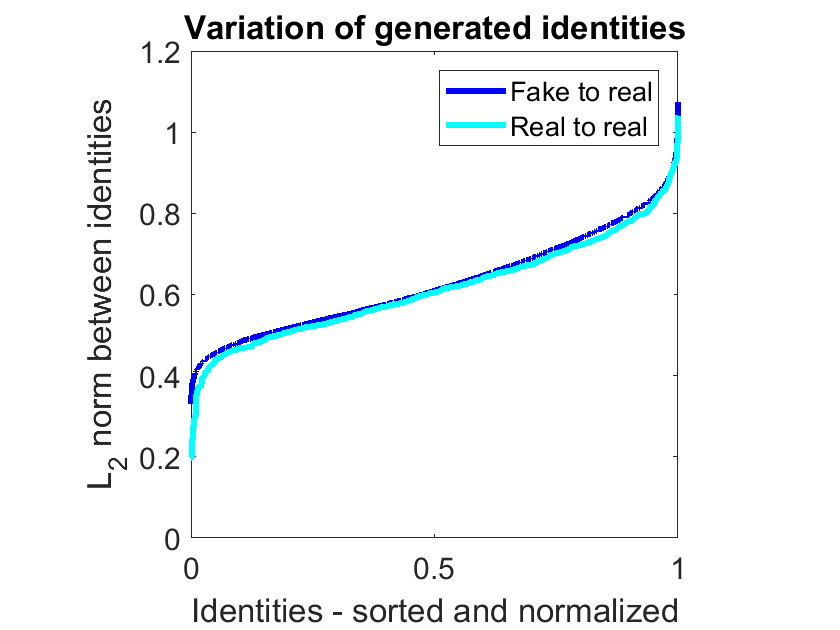}
    \caption{}
    \label{fig:variation_1}
    \end{subfigure}
    \hspace{1cm}
    \begin{subfigure}[t]{0.35\linewidth}
    \includegraphics[width = 1\linewidth,trim={2cm 0cm 2cm 0cm}, clip]{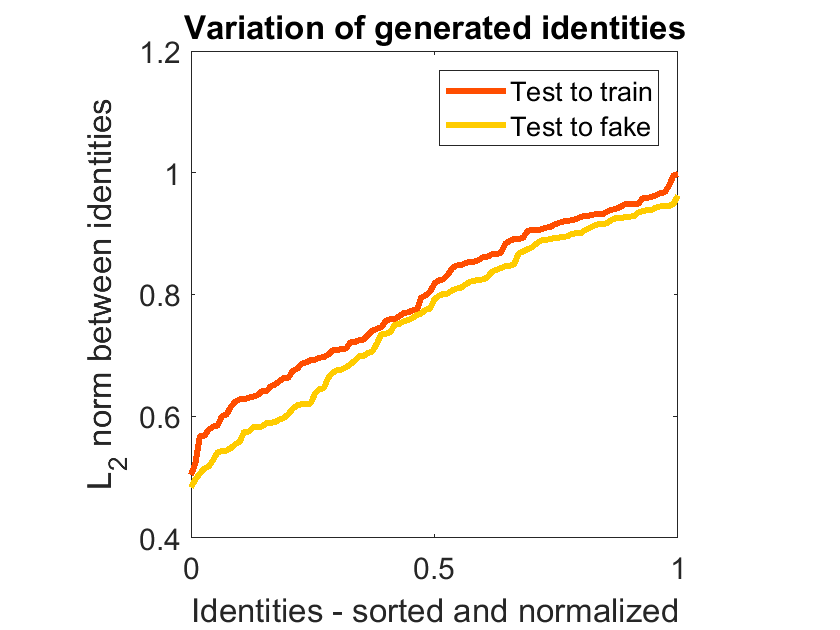}
    \caption{}
    \label{fig:variation_2}
    \end{subfigure}
    \caption{\small
    Distance between generated and real identities, measured as $L_2$ distance between identity descriptors. For all plots, the X axis is normalized and ordered by the distances.\\
    \textbf{(a) Fake to real:} for each generated identity, its distance to the nearest real training identity.
    \textbf{(a) Real to real:} for each real training identity, its distance to the nearest real training identity, excluding itself.
    \textbf{(b) Test to fake:} for each real test identity, its distance to the nearest generated identity.
    \textbf{(b) Test to train:} for each real test identity, its distance to the nearest real training identity.
    }
    \label{fig:variations}
\end{figure}

In the following experiment, we set out to demonstrate that our model is capable of generating novel identities and not just add small variations to the existing training data.
To that end, we made use of the identity descriptors extracted previously. We measured the $L_2$ distance between each generated identity and its closest real identity from the training data, and plotted the ordered distances. We repeated the process within the training data, namely, for each training identity, we measured the $L_2$ distance to its nearest neighbor within the training set (excluding itself).
\autoref{fig:variation_1} compares the resulting distances on a normalized axis.
It can be seen that the distributions of distances are similar, and that the generated faces are not bounded to small variations in the vicinity of the training samples, in terms of identity. In other words, the variation between generated samples and the existing data is at least as large as the variation of the data itself.

In order to test the ability of our model to generalize to a previously unseen test set of real faces, we held out roughly $5\%$ of the identities during training for evaluation.
We measured the $L_2$ distance between each real test set identity to the closest identity generated by the GAN, as well as to the closest real training set identity.
\autoref{fig:variation_2} compares the resulting ordered distances on a normalized axis.
It can be seen that the test set identities are closer to the generated identities than those of the training set. Moreover, the "Test to fake" distances are not significantly larger than the "Fake to real" distances in \autoref{fig:variation_1}. This implies that our model is capable of generating samples that diverge significantly from the original training set and may resemble previously unseen data.

\begin{figure}
\centering
\includegraphics[width=0.15\linewidth]{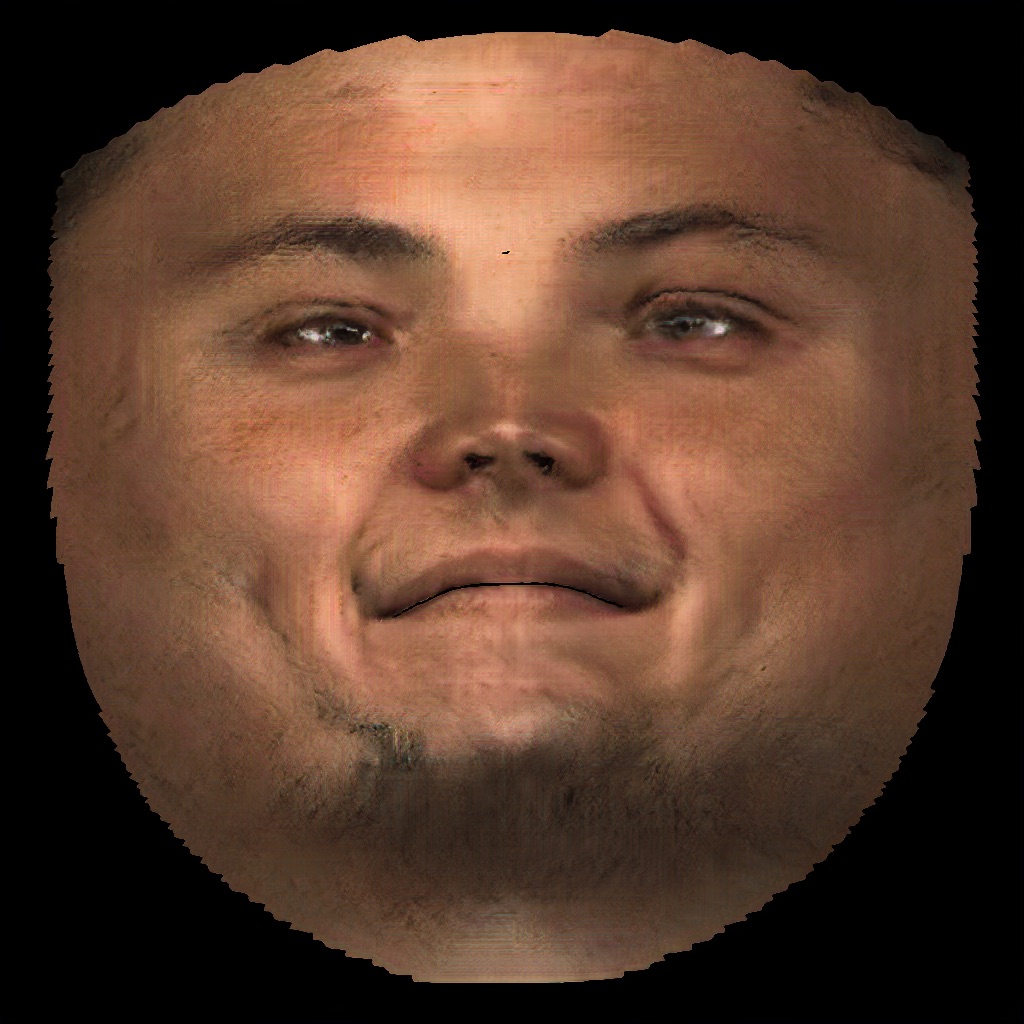}
\includegraphics[width=0.15\linewidth]{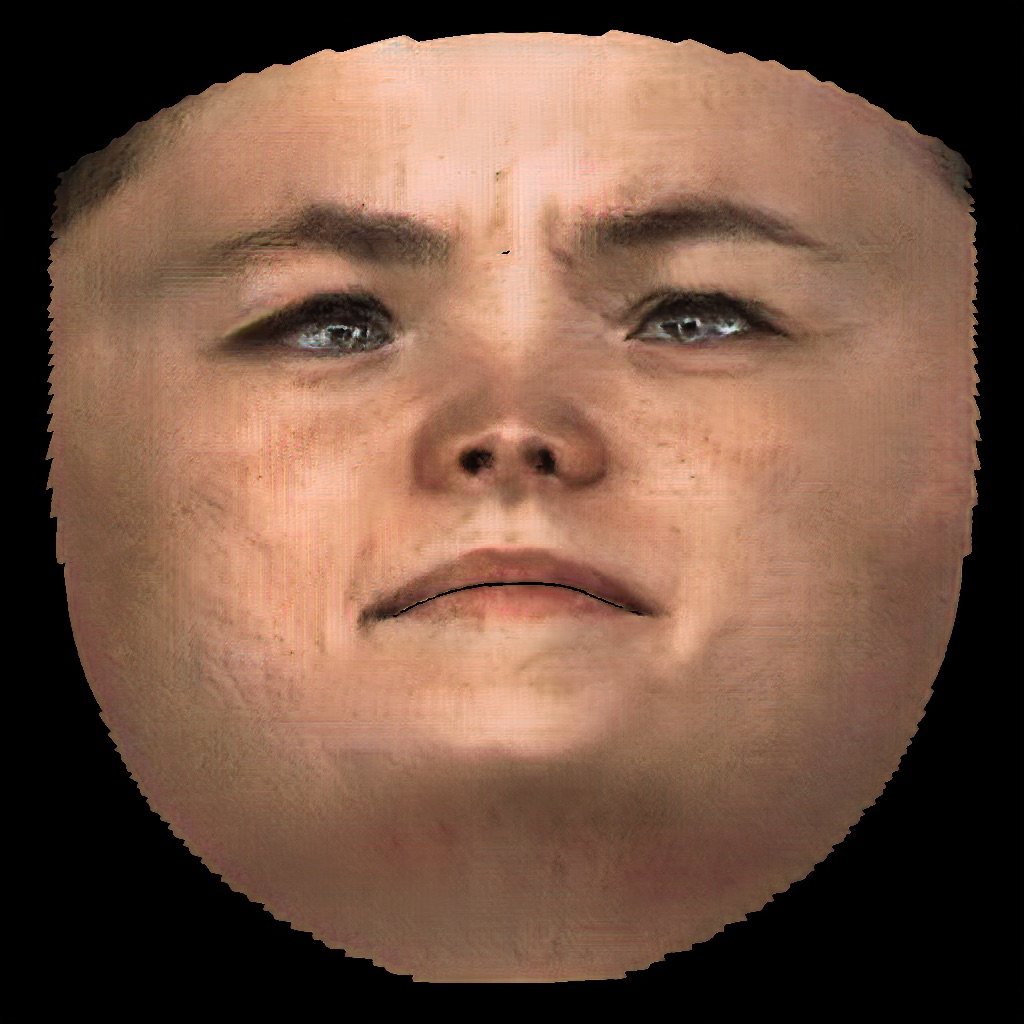}
\includegraphics[width=0.15\linewidth]{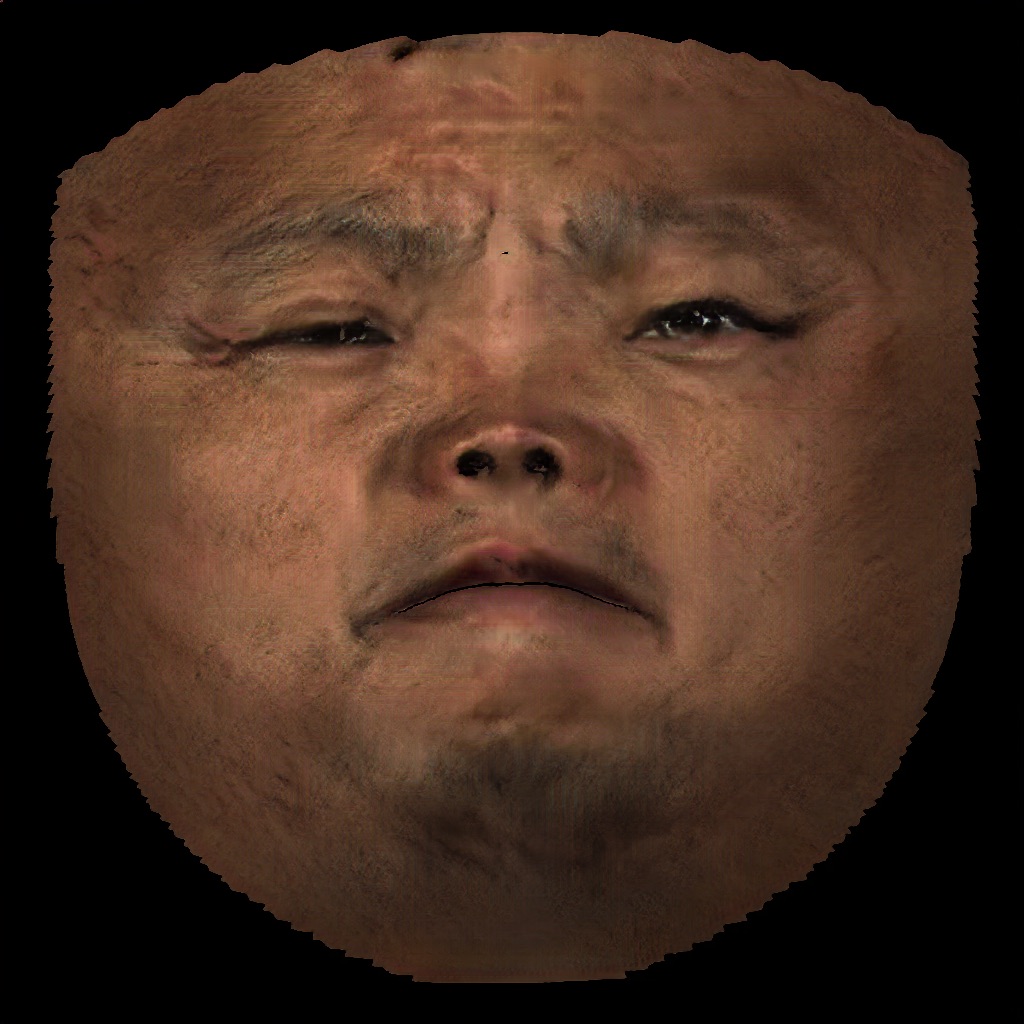}
\includegraphics[width=0.15\linewidth]{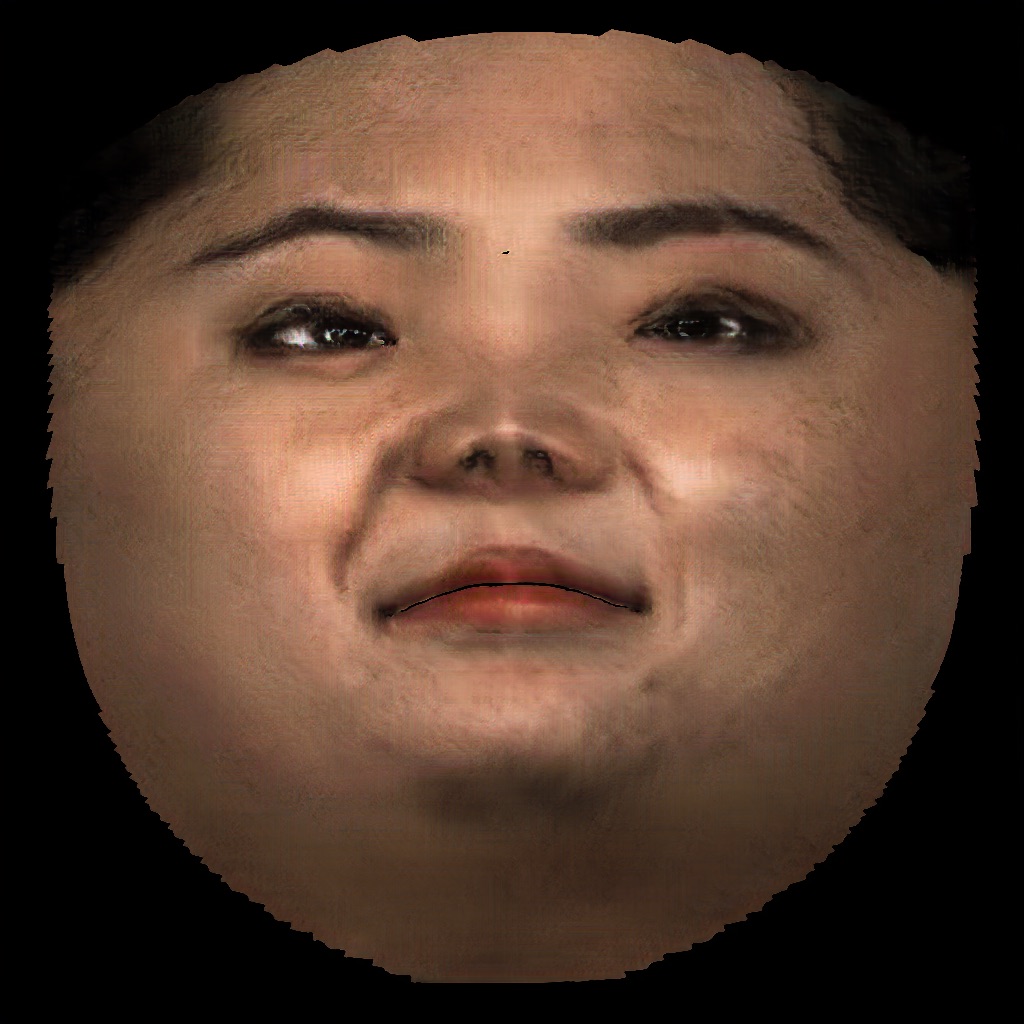}
\includegraphics[width=0.15\linewidth]{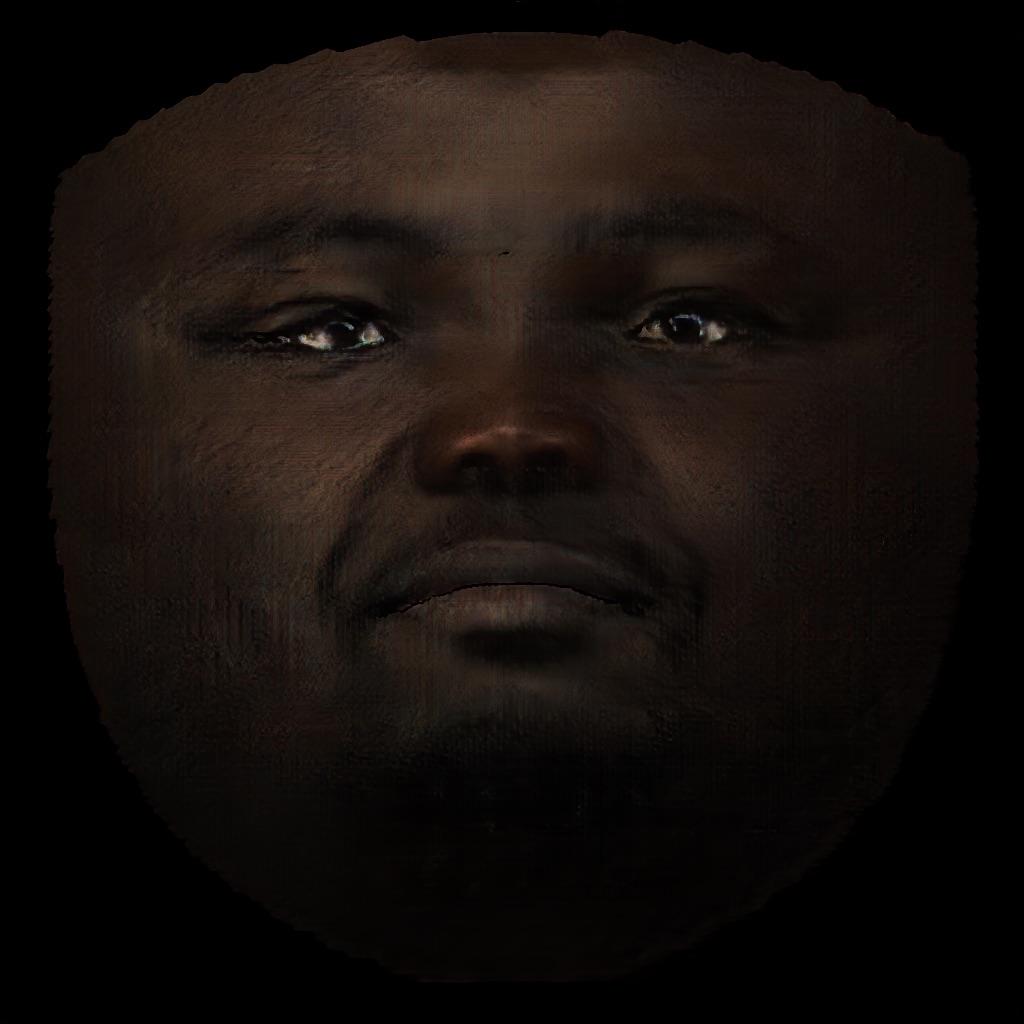}\\
\includegraphics[width=0.15\linewidth]{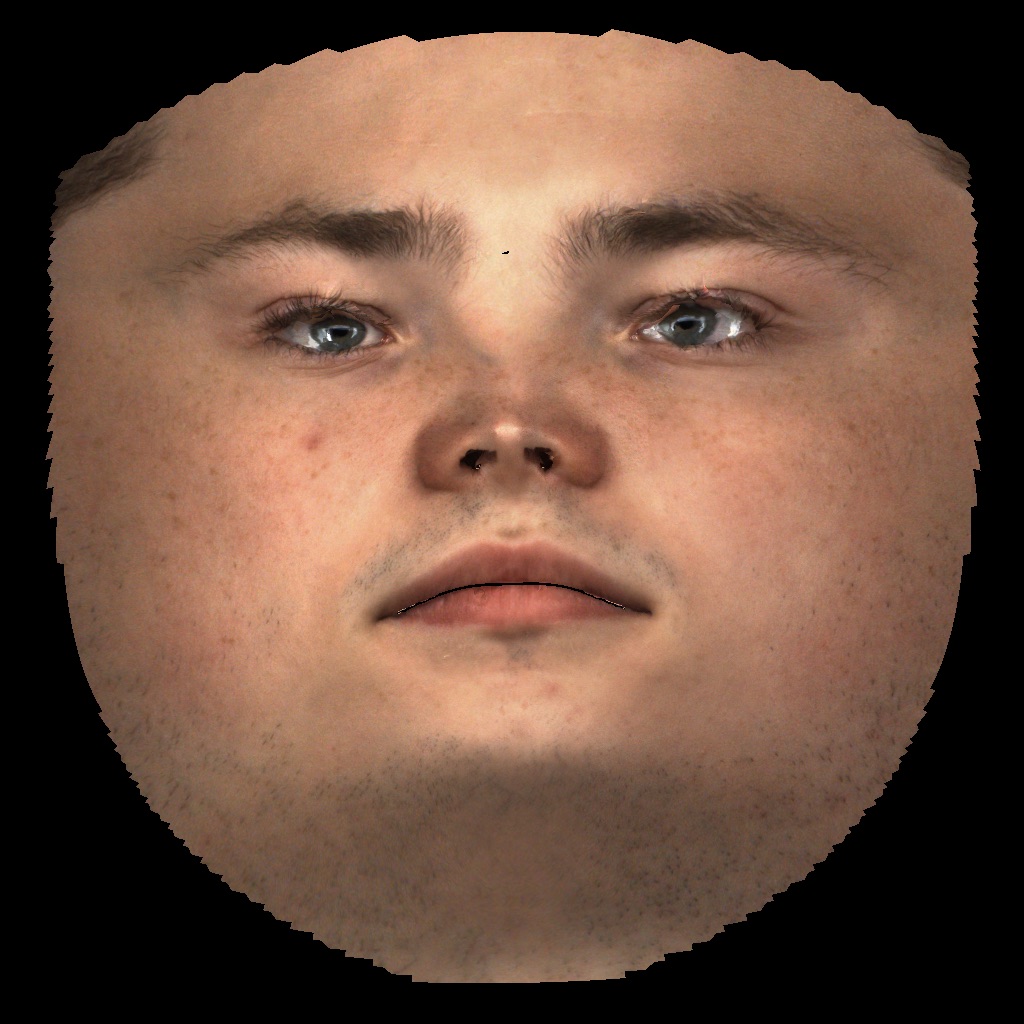}
\includegraphics[width=0.15\linewidth]{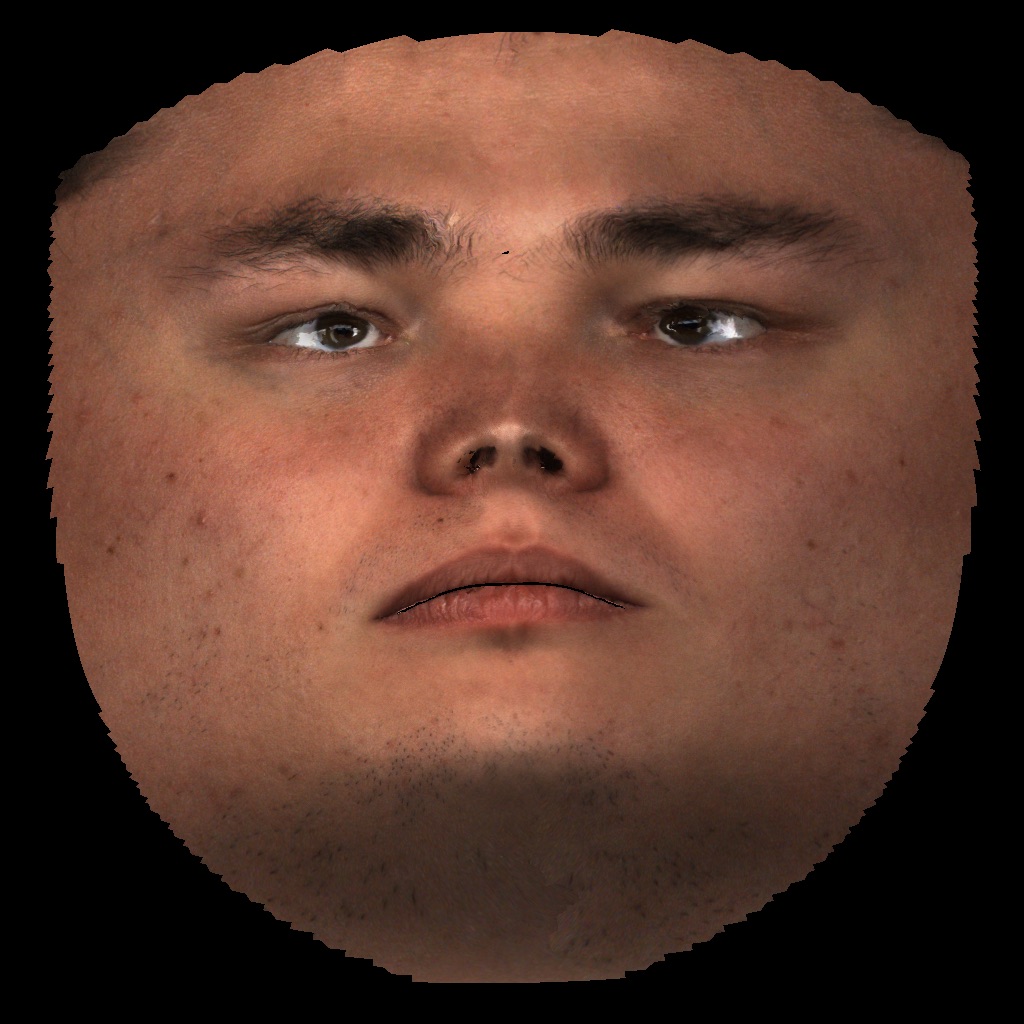}
\includegraphics[width=0.15\linewidth]{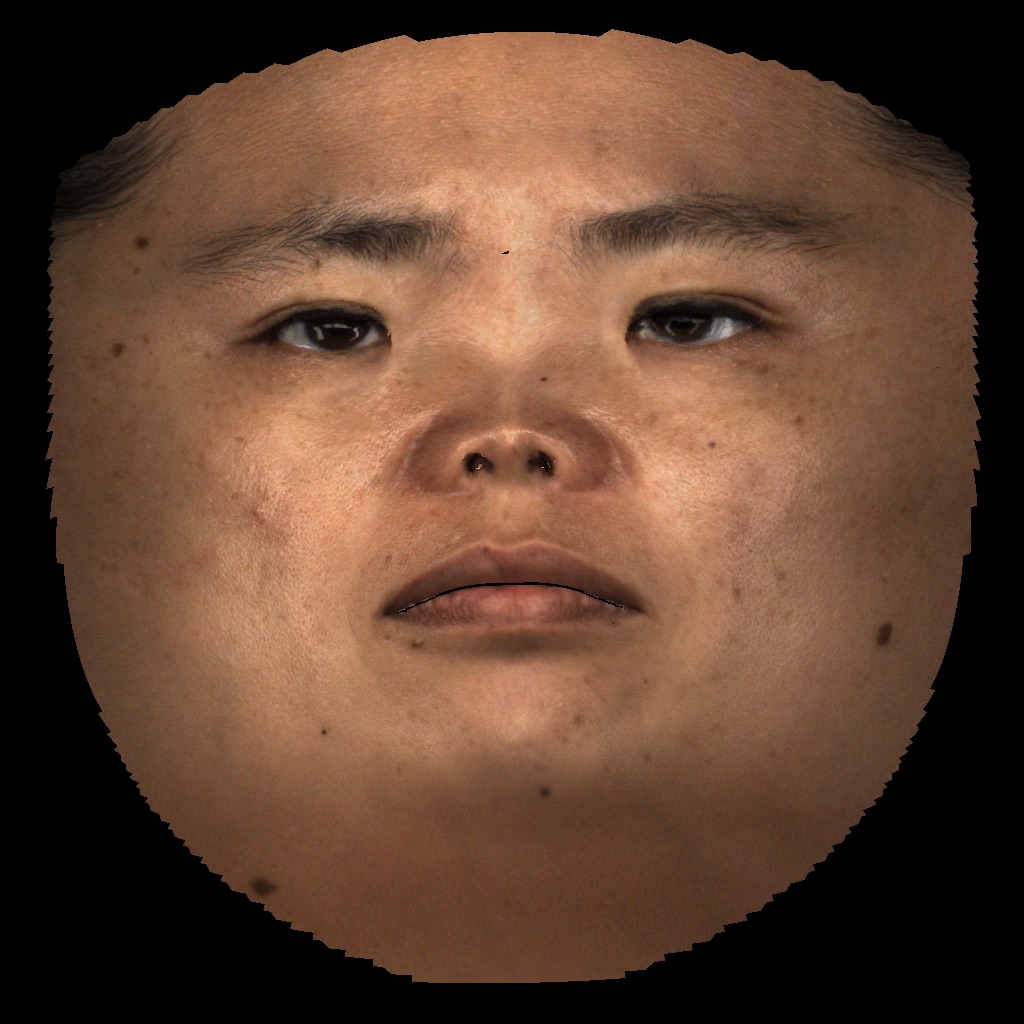}
\includegraphics[width=0.15\linewidth]{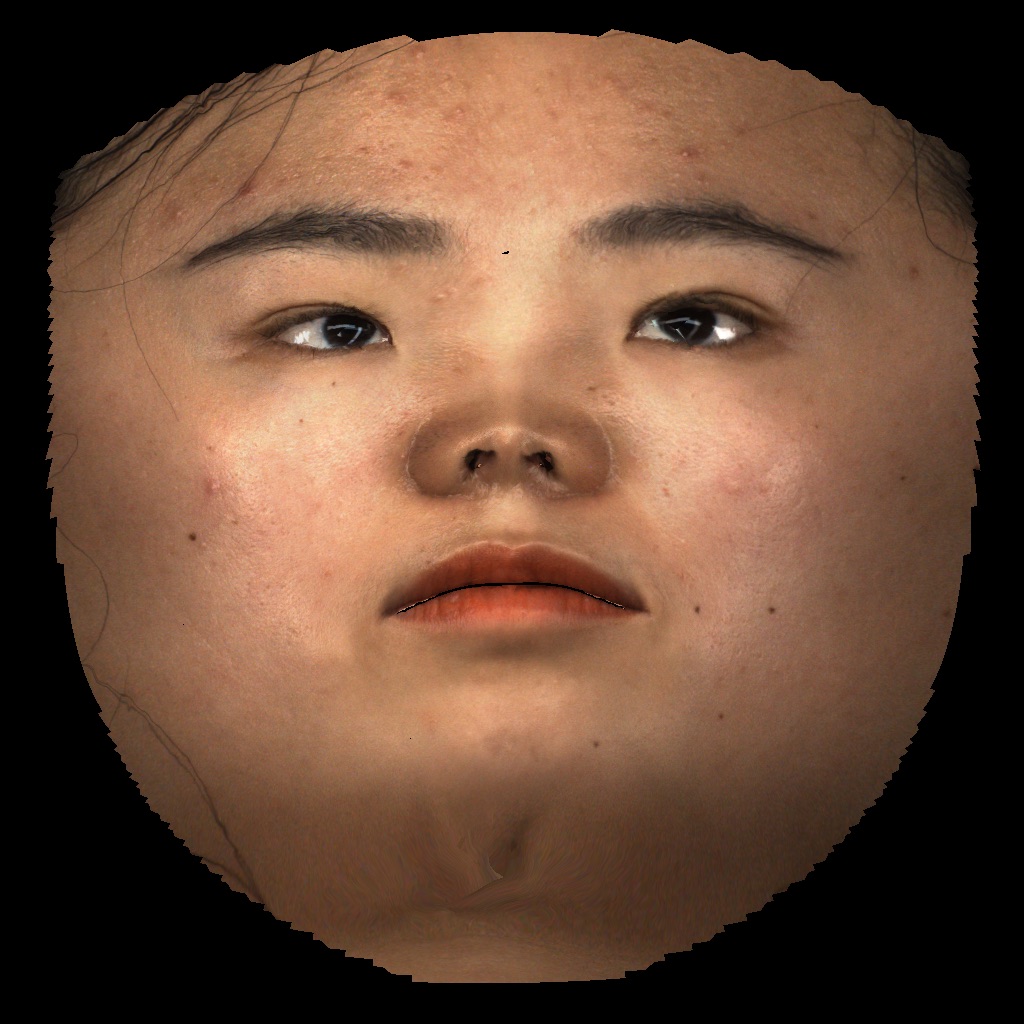}
\includegraphics[width=0.15\linewidth]{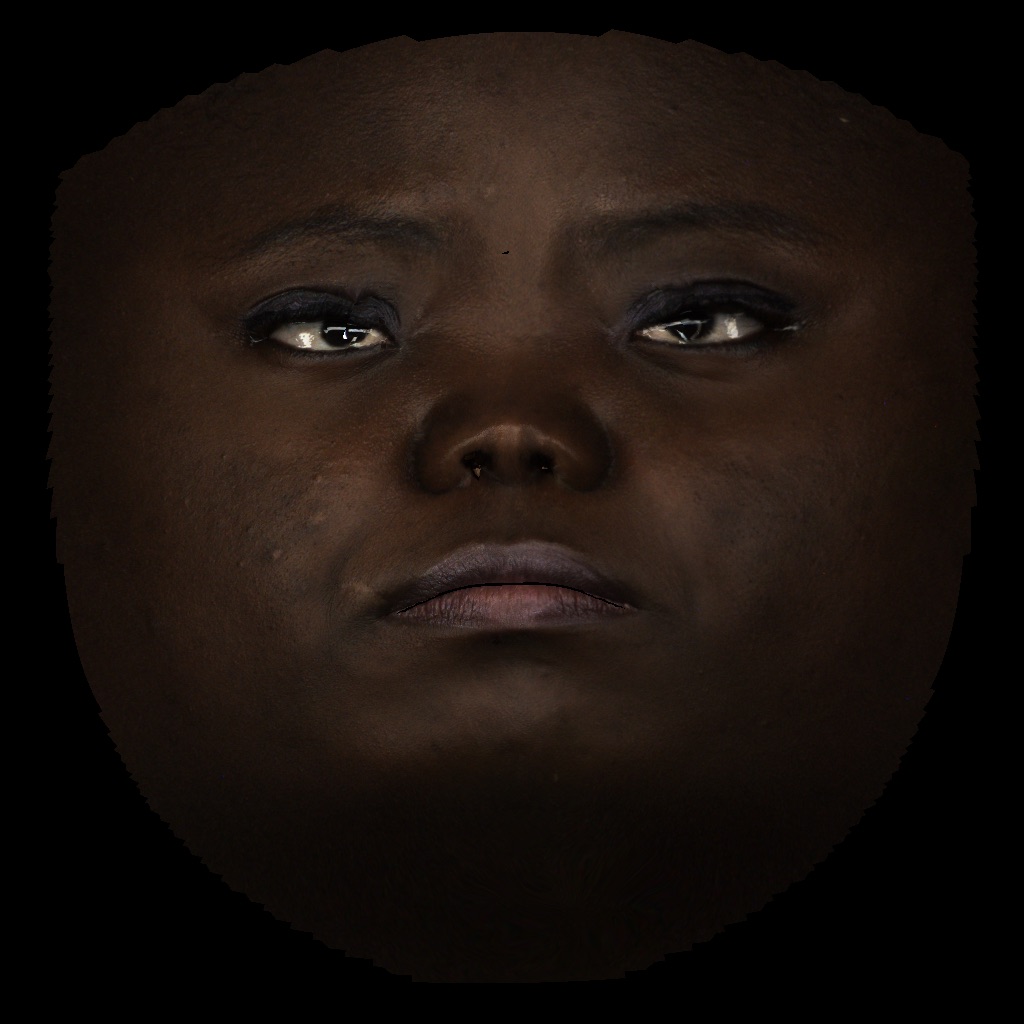}
\caption{Top: Synthesized facial textures. Bottom: corresponding closest real neighbors in terms of facial identity.}
\label{fig:NN}
\end{figure}

Finally, we performed a qualitative evaluation of the ability of our pipeline to generate original data samples. In \autoref{fig:NN} we show five textures generated by our proposed model, alongside the closest neighbor within the real data in sense of $L2$ norm between identity descriptors. This experiment indicates that the nearest real textures are far enough to be visually distinguished as different people, showing that our model is able to produce novel identities.

Extended results and illustrations are supplied within the supplementary material.

\section{Discussion and Conclusions}
\label{sec:discussion}
In this paper we present a new approach for synthesis of realistic 3D human faces. The proposed model, in contrast to 3DMM \cite{blanz1999morphable} is not limited to linear operations, and is able to exhibit more complex relations between the coefficients and the generated samples. In \autoref{sec:3dmm} we show the limitations of the 3DMM model in lack of complexity, realism and ability to correctly sample the true distribution. Namely, its simplified model follows a distribution that could generate non plausible samples. Although 3DMM might be capable of approximating real samples effectively, generating new plausible samples requires the model to conform with the correct data distribution prior.

We base our model on the notion of bringing the dataset into a uniform parametrization space, which allows for easier processing of geometry as images. This step allows to introduce the powerful tools of image processing into our geometric problem such as NN and more specifically, GANs. We use the GAN model in order to closely resemble the distribution of faces in our dataset, which allows to generate new samples that are both plausible and realistic.

The formation of geometries is performed per texture by learning the relation between texture and geometry coefficients of the 3DMM from the real data, following the observation that reducing the resolution of the geometry has negligible effect on the final appearance of the face. In \autoref{sec:geometry_gen}, we explored several methods for this purpose. In our experiments we used the LS method due to its low distortion and simplicity. The resulting geometries take into account expression, gender and race that appear in the texture, making the final result more realistic.

In \autoref{sec:expermintal} we preform several qualitative and quantitative evaluations in order to strengthen our claims. \autoref{fig:TSNE_experiment} depicts an embedding of real versus generated faces ID's, which demonstrates the ability of our proposed model to span the distribution of identities presented in the real data. It is important to note that the identities cover the same areas while filling the gaps between the real data samples. We also show that the 3DMM model sample distribution unsurprisingly resembles a Gaussian distribution which diverges from the training data distribution, although it was constructed based on this data.

Further experimentation depicted in \autoref{fig:variations} shows results of nearest neighbor searches between generated and real samples. It is important to note that \autoref{fig:variation_2} depicts the relation between real and generated samples to test set samples which were held out during training. We were able to show that the distances of real samples to the test set tend to be higher than distances from the generated samples to the same test set. This demonstrates that generated samples do not just produce IDs that are very close to the training set, but also novel IDs that resemble previously unseen examples.

We believe that this general framework for modeling geometry and texture can be useful for many applications. One prominent example is to use our proposed model in order synthesize more realistic facial data which can be used to train face detection, face recognition or face reconstruction models. We also believe that our model can be valuable in cases where many different realistic faces need to be created, such as in film industry or computer games.
Note that this does not require to generalize the training samples but only to to produce various different plausible facial samples.
This general methodology can also be employed for different various classes of objects where alignment of the data is possible.

\clearpage

\bibliographystyle{splncs}
\bibliography{egbib}

\begin{thebibliography}{10}

\bibitem{masi2016we}
Masi, I., Trần, A.T., Hassner, T., Leksut, J.T., Medioni, G.:
\newblock Do we really need to collect millions of faces for effective face
  recognition?
\newblock In: European Conference on Computer Vision, Springer (2016)  579--596

\bibitem{shrivastava2017learning}
Shrivastava, A., Pfister, T., Tuzel, O., Susskind, J., Wang, W., Webb, R.:
\newblock Learning from simulated and unsupervised images through adversarial
  training.
\newblock In: The IEEE Conference on Computer Vision and Pattern Recognition
  (CVPR). Volume~3. (2017) ~6

\bibitem{richardson2017learning}
Richardson, E., Sela, M., Or-El, R., Kimmel, R.:
\newblock Learning detailed face reconstruction from a single image.
\newblock In: 2017 IEEE Conference on Computer Vision and Pattern Recognition
  (CVPR), IEEE (2017)  5553--5562

\bibitem{gecer2018semi}
Gecer, B., Bhattarai, B., Kittler, J., Kim, T.K.:
\newblock Semi-supervised adversarial learning to generate photorealistic face
  images of new identities from 3d morphable model.
\newblock arXiv preprint arXiv:1804.03675 (2018)

\bibitem{blanz1999morphable}
Blanz, V., Vetter, T.:
\newblock A morphable model for the synthesis of 3d faces.
\newblock In: Proceedings of the 26th annual conference on Computer graphics
  and interactive techniques, ACM Press/Addison-Wesley Publishing Co. (1999)
  187--194

\bibitem{saito2017photorealistic}
Saito, S., Wei, L., Hu, L., Nagano, K., Li, H.:
\newblock Photorealistic facial texture inference using deep neural networks.
\newblock In: IEEE Conference on Computer Vision and Pattern Recognition, CVPR.
  Volume~3. (2017)

\bibitem{richardson20163d}
Richardson, E., Sela, M., Kimmel, R.:
\newblock 3d face reconstruction by learning from synthetic data.
\newblock In: 3D Vision (3DV), 2016 Fourth International Conference on, IEEE
  (2016)  460--469

\bibitem{sela2017unrestricted}
Sela, M., Richardson, E., Kimmel, R.:
\newblock Unrestricted facial geometry reconstruction using image-to-image
  translation.
\newblock In: 2017 IEEE International Conference on Computer Vision (ICCV),
  IEEE (2017)  1585--1594

\bibitem{jolliffe1986principal}
Jolliffe, I.T.:
\newblock Principal component analysis and factor analysis.
\newblock In: Principal component analysis.
\newblock Springer (1986)  115--128

\bibitem{booth2018large}
Booth, J., Roussos, A., Ponniah, A., Dunaway, D., Zafeiriou, S.:
\newblock Large scale 3d morphable models.
\newblock International Journal of Computer Vision \textbf{126}(2-4) (2018)
  233--254

\bibitem{booth20163d}
Booth, J., Roussos, A., Zafeiriou, S., Ponniah, A., Dunaway, D.:
\newblock A 3d morphable model learnt from 10,000 faces.
\newblock In: Proceedings of the IEEE Conference on Computer Vision and Pattern
  Recognition. (2016)  5543--5552

\bibitem{gross2010multi}
Gross, R., Matthews, I., Cohn, J., Kanade, T., Baker, S.:
\newblock Multi-pie.
\newblock Image and Vision Computing \textbf{28}(5) (2010)  807--813

\bibitem{menpo14}
{Alabort-i-Medina}, J., Antonakos, E., Booth, J., Snape, P., Zafeiriou, S.:
\newblock Menpo: A comprehensive platform for parametric image alignment and
  visual deformable models.
\newblock In: Proceedings of the ACM International Conference on Multimedia. MM
  '14, New York, NY, USA, ACM (2014)  679--682

\bibitem{dlib09}
King, D.E.:
\newblock Dlib-ml: A machine learning toolkit.
\newblock Journal of Machine Learning Research \textbf{10} (2009)  1755--1758

\bibitem{blender}
{Blender Online Community}:
\newblock Blender - a 3D modelling and rendering package.
\newblock Blender Foundation, Blender Institute, Amsterdam. (2017)
  \url{http://www.blender.org}.

\bibitem{goodfellow2014generative}
Goodfellow, I., Pouget-Abadie, J., Mirza, M., Xu, B., Warde-Farley, D., Ozair,
  S., Courville, A., Bengio, Y.:
\newblock Generative adversarial nets.
\newblock In: Advances in neural information processing systems. (2014)
  2672--2680

\bibitem{oord2016wavenet}
Oord, A.v.d., Dieleman, S., Zen, H., Simonyan, K., Vinyals, O., Graves, A.,
  Kalchbrenner, N., Senior, A., Kavukcuoglu, K.:
\newblock Wavenet: A generative model for raw audio.
\newblock arXiv preprint arXiv:1609.03499 (2016)

\bibitem{karras2017progressive}
Karras, T., Aila, T., Laine, S., Lehtinen, J.:
\newblock Progressive growing of gans for improved quality, stability, and
  variation.
\newblock nternational Conference on Learning Representations (ICLR) (2017)

\bibitem{isola2017image}
Isola, P., Zhu, J.Y., Zhou, T., Efros, A.A.:
\newblock Image-to-image translation with conditional adversarial networks.
\newblock arXiv preprint (2017)

\bibitem{zhu2017unpaired}
Zhu, J.Y., Park, T., Isola, P., Efros, A.A.:
\newblock Unpaired image-to-image translation using cycle-consistent
  adversarial networks.
\newblock arXiv preprint arXiv:1703.10593 (2017)

\bibitem{gulrajani2017improved}
Gulrajani, I., Ahmed, F., Arjovsky, M., Dumoulin, V., Courville, A.C.:
\newblock Improved training of wasserstein gans.
\newblock In: Advances in Neural Information Processing Systems. (2017)
  5769--5779

\bibitem{arjovsky2017wasserstein}
Arjovsky, M., Chintala, S., Bottou, L.:
\newblock Wasserstein gan.
\newblock arXiv preprint arXiv:1701.07875 (2017)

\bibitem{mao2017least}
Mao, X., Li, Q., Xie, H., Lau, R.Y., Wang, Z., Smolley, S.P.:
\newblock Least squares generative adversarial networks.
\newblock In: 2017 IEEE International Conference on Computer Vision (ICCV),
  IEEE (2017)  2813--2821

\bibitem{chu20143d}
Chu, B., Romdhani, S., Chen, L.:
\newblock 3d-aided face recognition robust to expression and pose variations.
\newblock In: Proceedings of the IEEE Conference on Computer Vision and Pattern
  Recognition. (2014)  1899--1906

\bibitem{rabin2011wasserstein}
Rabin, J., Peyr{\'e}, G., Delon, J., Bernot, M.:
\newblock Wasserstein barycenter and its application to texture mixing.
\newblock In: International Conference on Scale Space and Variational Methods
  in Computer Vision, Springer (2011)  435--446

\bibitem{maaten2008visualizing}
Maaten, L.v.d., Hinton, G.:
\newblock Visualizing data using t-sne.
\newblock Journal of machine learning research \textbf{9}(Nov) (2008)
  2579--2605

\bibitem{amos2016openface}
Amos, B., Ludwiczuk, B., Satyanarayanan, M.:
\newblock Openface: A general-purpose face recognition library with mobile
  applications.
\newblock Technical report, CMU-CS-16-118, CMU School of Computer Science
  (2016)

\end{thebibliography}
\clearpage

\section*{Supplemantary material}
\subsection*{Maximum likelihood approach}
In order to correlate between geometry and texture, we construct a joint 3DMM by concatenating the texture and geometry vectors to each other. Denote by $M=
\begin{pmatrix}
G\\T
\end{pmatrix}$
 the $6m \times n$ matrix which consists of the geometries $G$ and textures $T$ vertically concatenated, and define $\Delta M = M - \mu_M\mathbbm{1}^T$.
 Denote by $U$ the matrix that contains the basis vectors of $\Delta M$ in its columns.
 $U$ can be computed either by the eigenvalue decomposition of $\Delta M \Delta M^T$ as the eigenvectors, or more efficiently by the singular value decomposition of $\Delta M$ as the left singular vectors.
 We assume that the columns of $U$ are ordered by the size of their corresponding singular values in a descending manner.
 Denote by $U_g$ and $U_t$ the upper and lower halves of $U$, such that $U=
\begin{pmatrix}
U_g\\U_t
\end{pmatrix}$.
Similarily, denote $\mu_M =
\begin{pmatrix}
\mu_{M_g} \\ \mu_{M_t}.
\end{pmatrix}$.
Note that $U_g$ and $U_t$ are not equivalent to the previously defined $V_g$ and $V_t$ and are not orthogonal. Still, any geometry $g$ and texture $t$ of a given face in $M$ can be represented as a linear combination
\begin{equation}
    \begin{pmatrix} g \\ t \end{pmatrix} =
    \begin{pmatrix} \mu_{M_g} \\ \mu_{M_t} \end{pmatrix} +
    \begin{pmatrix} U_g \\ U_t \end{pmatrix}\beta,
\end{equation}
where the coefficient vector $\beta$ is mutual to the geometry and texture.

Given a coefficient vector $\beta$ of a new face that was not used to construct the model, one can formulate the texture of the face as
\begin{equation}
    t = U_t\beta + \mu_{M_t} + noise
    \label{eq:texture_noise_model}
\end{equation}
Our goal is to estimate $\beta$ given a texture $t$. According to the maximum likelihood approach, $\beta$ is estimated as
\begin{eqnarray}
\beta^* &=& \argmax_{\beta} P(\beta|t) \cr
&=& \argmax_{\beta} P(t|\beta)P(\beta).
\end{eqnarray}
Assuming, as before, that $P(t|\beta)$ and $P(\beta)$ follow a multivariate normal distribution with diagonal covariance matrices $\Sigma_{t|\beta}, \Sigma_{\beta}$ and mean $\mu_{t|\beta} = U_t\beta,\ \ \  \mu_\beta = \vec{0}.$
%
Then, one can write
\begin{eqnarray}
\beta^* &=& \argmax_\beta \exp\left \{-\frac{1}{2}(t-\mu_{t|\beta})\Sigma_{t|\beta}^{-1}(t-\mu_{t|\beta})^T\right \}\cdot \cr
&& \exp\left \{-\frac{1}{2}\beta\Sigma_{\beta}^{-1}\beta^T\right \} \cr
&=& \argmin_{\beta} (t-U_t\beta)\Sigma_{t|\beta}^{-1}(t-V_t\beta)^T + \beta\Sigma_{\beta}^{-1}\beta^T \,\,\,.
\end{eqnarray}
The solution $\beta^*$ is obtained by taking the gradient to zero, which yields
\begin{equation}
\beta^* = (U_t^T\Sigma_{t|\beta}^{-1}U_t+\Sigma_{\beta})^{-1}(t^T\Sigma_{t|\beta}^{-1}U_t).
\end{equation}
We estimate the covariance matrix $\Sigma_{\beta}$ empirically from the data. The covariance matrix $\Sigma_{t|\beta}$ can be estimated as well, however an arbitrary scale parameter which determines strength of the prior  $P(\beta)$ relative to the data must be calibrated.
In other words, this scale takes into account the noise level described in  \autoref{eq:texture_noise_model}.
Finally, once the coefficient vector $\beta$ is known, the geometry is obtained by $g = U_g\beta.$
\subsection*{Blend Shapes for different expressions}

\begin{figure*}[!h]
\centering
\includegraphics[width=1\linewidth]{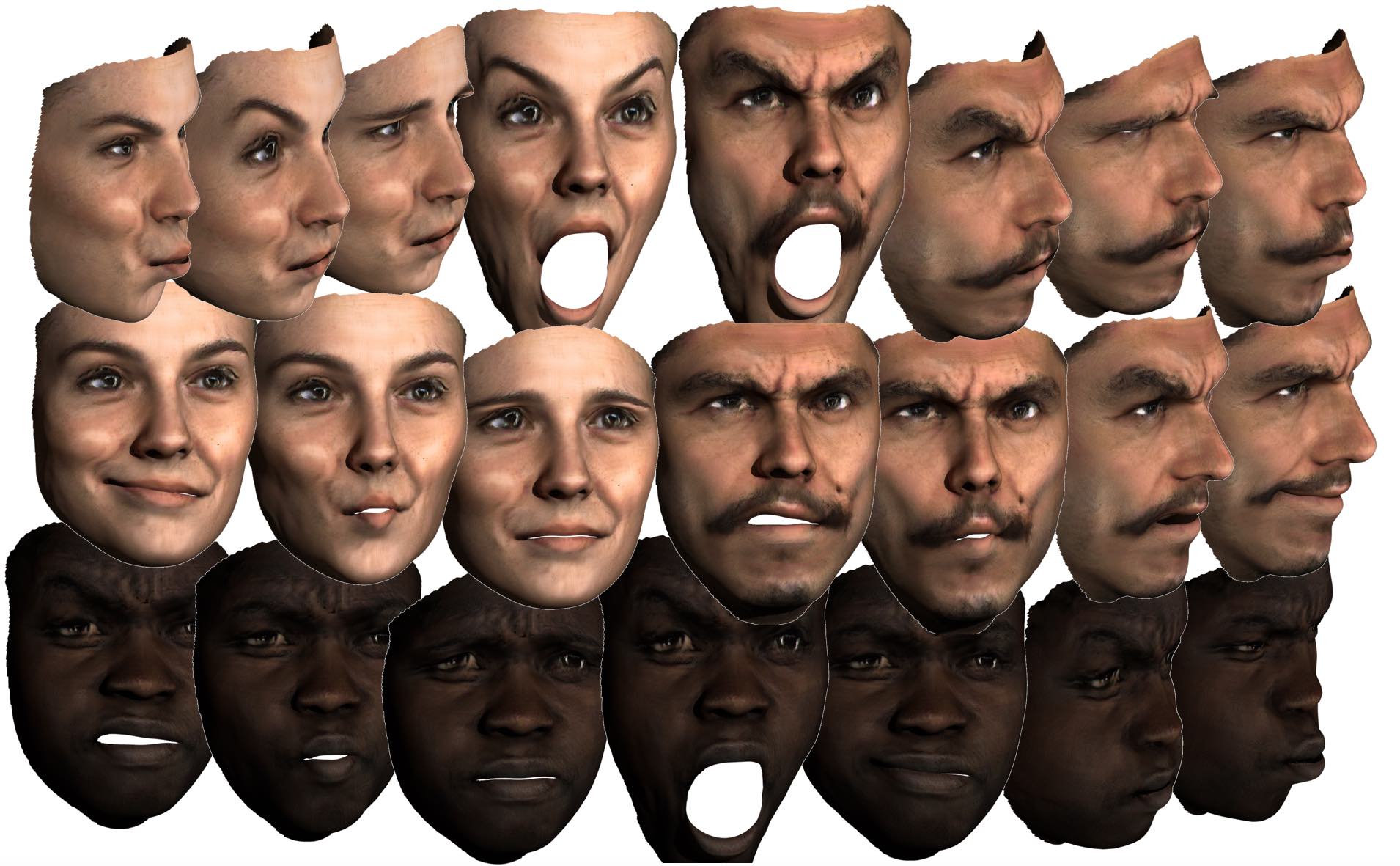}
\caption{Three identities generated by the proposed method, deformed with different Blend Shapes expressions.}
\end{figure*}

\clearpage
\subsection*{Different poses and lighting conditions}
\begin{figure*}[!h]
\centering
\includegraphics[width=1\linewidth]{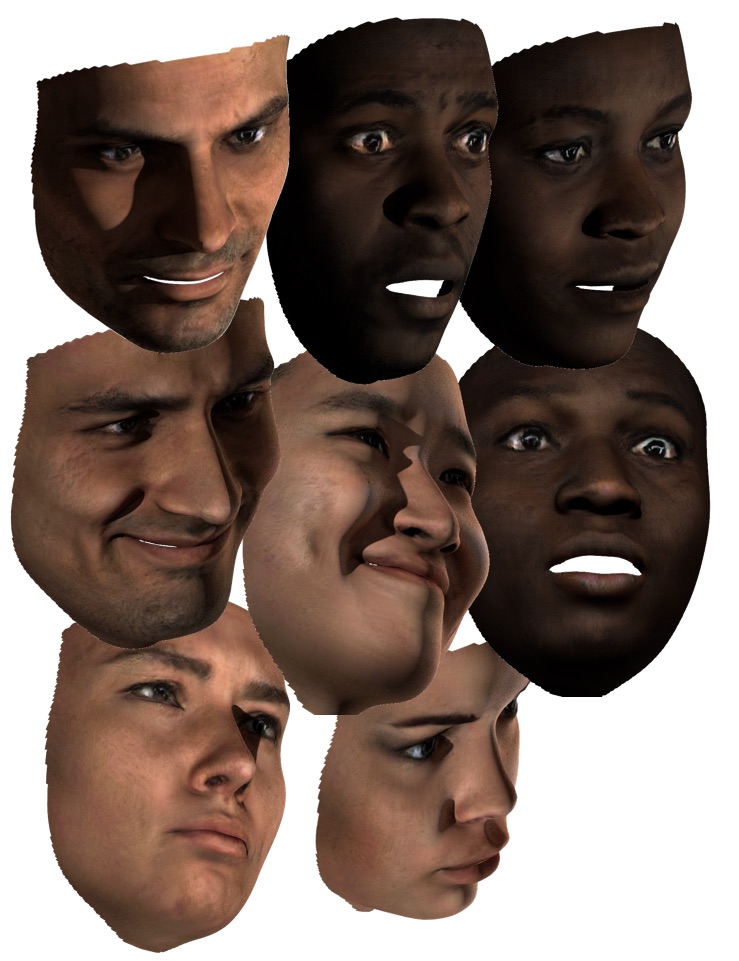}
\caption{Identities generated by the proposed method with different pose and lighting.}
\end{figure*}

\begin{figure*}[!h]
\centering
\includegraphics[width=1\linewidth]{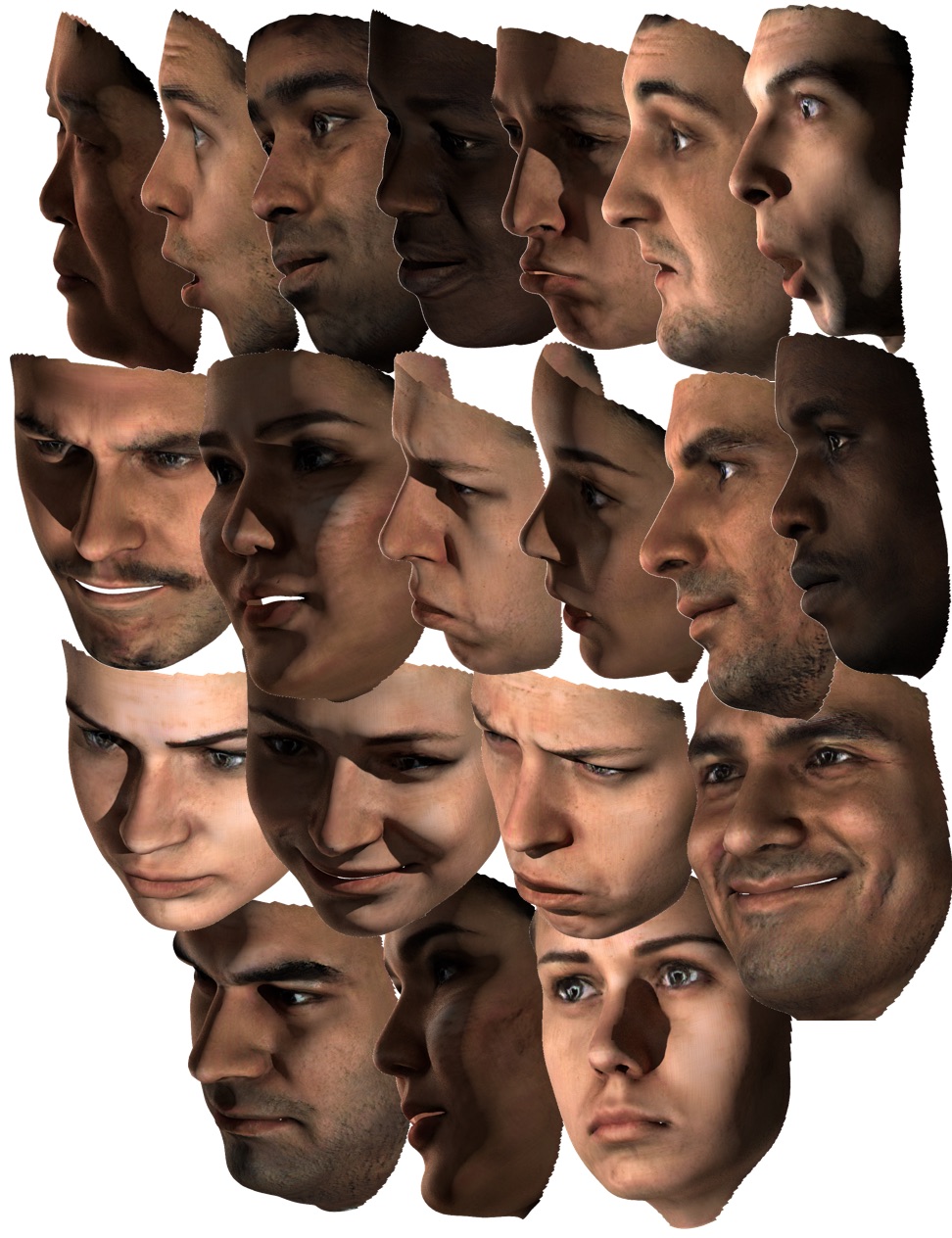}
\caption{Identities generated by the proposed method with different pose and lighting.}
\end{figure*}

\end{document}